\documentclass[fleqn,usenatbib]{mnras}

\usepackage{newtxtext,newtxmath}
\usepackage[T1]{fontenc}

\DeclareRobustCommand{\VAN}[3]{#2}
\let\VANthebibliography\thebibliography
\def\thebibliography{\DeclareRobustCommand{\VAN}[3]{##3}\VANthebibliography}

\usepackage[hang,small,bf]{caption}
\usepackage[subrefformat=parens]{subcaption}
\usepackage{multicol}   %
\usepackage{multirow}

\usepackage{acronym}

\usepackage{graphicx}	% Including figure files
\usepackage{amsmath}	% Advanced maths commands
\usepackage{newtxtext,newtxmath}
\newcommand{\bm}[1]{{\mbox{\boldmath $#1$}}}

\acrodef{EOS}[EOS]{equation of state}
\acrodef{WD}[WD]{white dwarf}
\acrodefplural{WD}[WDs]{white dwarfs}
\acrodef{MS}[MS]{main sequence}

\title[A Lagrangian construction of rotating stars]{A Lagrangian construction of rotating stars}

\author[M. Ogata et al.]{
Misa Ogata,$^{1}$\thanks{E-mail: ogata@heap.phys.waseda.ac.jp}
Hirotada Okawa, $^{2}$
Kotaro Fujisawa, $^{3}$
Nobutoshi Yasutake,$^{4,5}$
Yu Yamamoto,$^{1}$
\newauthor{and Shoichi Yamada $^{1}$}
\\
$^{1}$ Research Institute for Science and Engineering, Waseda University, Tokyo 169-8555, Japan\\
$^{2}$ Waseda Institute for Advanced Study (WIAS), 1-21-1 Nishi Waseda, Shinjuku, Tokyo 169-0051, Japan \\
$^{3}$ Department of Physics, Graduate School of Science, the University of Tokyo, Bunkyo-ku, Tokyo 113-0033, Japan \\
$^{4}$ Physics Department, Chiba Institute of Technology, Chiba 275-0023, Japan \\
$^{5}$ Advanced Science Research Center, Japan Atomic Energy Agency, Tokai, Ibaraki 319-1195, Japan  \\
}

\date{Accepted XXX. Received YYY; in original form ZZZ}

\pubyear{2022}

\begin{document}
\label{firstpage}
\pagerange{\pageref{firstpage}--\pageref{lastpage}}
\maketitle

\begin{abstract}
We present a new formulation for numerically obtaining axisymmetric equilibrium structures of rotating stars in two spatial dimensions. With a view to apply it to the secular evolution of rotating stars, we base it on the Lagrangian description, i.e., we solve the force-balance equations to find the spatial positions of fluid elements endowed individually with a mass, specific entropy and angular momentum. The system of nonlinear equations obtained by finite-differencing the basic equations are solved with the W4 method, which is a new multi-dimensional root-finding scheme of our own devising. We augment it with a remapping scheme to avoid distortions of the Lagrangian coordinates. In this first one of a series of papers, we will give a detailed description of these methods initially. We then present the results of some test calculations, which include the construction of both rapidly rotating barotropic and baroclinic equilibrium states. We gauge their accuracies quantitatively with some diagnostic quantities as well as via comparisons with the counterparts obtained with an Eulerian code. For a demonstrative purpose, we apply the code to a toy-model cooling calculation of a rotating white dwarf.
\end{abstract}


%%%%%%%%%%%%%%%%%%%%%%%%%%%%%%%%%%%%%%%%%%%%%%%%%%

%%%%%%%%%%%%%%%%% BODY OF PAPER %%%%%%%%%%%%%%%%%%

\section{Introduction}

The evolution of rotating stars has been studied intensively and extensively over the years \citep[e.g.,][]{Maeder2000,Woosley2006}. The description is still incomplete, though, since those studies have commonly employed spatially one-dimensional (1D) models, in which the angular dependence of stellar structures is somehow averaged.
One of the major effects of rotation is the flattening of stars by centrifugal forces. There is observational evidence that the shapes of some rapidly-rotating stars are indeed significantly non-spherical \citep[][]{McAlister2005,Che2011}. The evolutions of such highly asymmetric stars may not be fully captured by the 1D models.

The effects of rotation are not limited to the stellar shape. The convection is affected by rotation, for instance. The ordinary (Schwarzschild or Ledoux) criterion should be replaced by the H\o iland criterion \citep{Tassoul1978}, in which stabilization by rotation is taken into account. There are other rotation-related instabilities, such as the Goldreich-Schubert-Fricke instabiliry \citep{Goldreich1967,fricke1968} and the ABCD instability \citep{Knobloch1983}, that may affect the matter composition, which should be also non-spherical, as well as the angular momentum distribution itself.
For example, \citet{Meynet2000} demonstrated that the rotationally-induced mixing can enhance the supply of hydrogen and extend the \ac{MS} lifetime by $\sim 30 \%$; they also showed that rotation leads to the He- and N-enhancement on the stellar surface. It is also mentioned that the mass loss should occur anisotropically.

It has been known observationally that intermediate- and high-mass stars are commonly rotating rapidly: some of them have surface rotational velocities as high as $300-400\ {\mathrm{km\ s^{-1}}}$ \citep[][]{Zorec2012,Ramirez-Agudelo2013};
Be stars \citep[][]{Porter2003,Rivinius2013}, which are surrounded by a disk, are found to rotate at about 75\% of the critical rotation velocity on average \citep[][]{Cochetti2019}.
Recently many researchers are attracted by merger events and their remnants: \citet{Schneider2019} studied the evolution of rapidly-rotating merger remnants of \ac{MS}-\ac{MS} binaries using a 1D evolutionary calculation code; \citet{Sun2021} calculated the evolution of a blue straggler, which appears younger than the real age as a result of the mass- and angular-momentum transfer from the companion in a close binary, also in 1D. Since there is no multi-dimensional stellar evolution code at the moment, we have no choice but to use such 1D evolutionary codes even when rotation is substantial.

The 1D stellar evolution codes have been very successful in understanding the evolution of various stars with different masses and metallicities, all essentially derived from the groundbreaking numerical method by \citet{Henyey1960}. They have been extended over the years by incorporating various mixing processes, rotation and magnetic field and so on \citep[e.g.][]{Meynet1997,Woosley2002}. However, as they are 1D, the multi-dimensional effects of rotation mentioned above are taken into account only in the average sense. In fact, the rotation law, for example, is mostly assumed to be shellular when one takes the angular average \citep[][]{Zahn1992,Meynet1997}.

Since the evolutionary timescale is longer than the dynamical timescale by many orders, stellar evolution calculations are based on hydrostatic equilibrium structures, on top of which nuclear burning and energy transport are computed.
In order to follow the secular evolution of stars in multi-dimensions, it is hence indispensable to obtain rotational equilibrium configurations numerically. In so doing, the Lagrangian formulation would certainly have an advantage over the Eulerian formulation if one were to employ it to the evolutionary calculation. As a matter of fact, the mass coordinate is always used in 1D calculations. This is because nuclear reactions occur locally in each fluid element and the resultant nuclear composition is carried with it as the star contracts or expands during its evolution. In the case of rotational stars, the specific angular momentum (in addition to the specific entropy) is another property of the fluid element: it is conserved along the stream line in the absence of angular momentum transfer among fluid elements. It would be very difficult to calculate such advection on the Eulerian coordinates that occurs very slowly on the secular timescale.
The formulation on the Lagrange coordinate in 1D is almost trivial: one has only to use the mass coordinate whereas in multi-dimensions it is highly nontrivial and is actually one of the major obstacles for the multi-dimensional calculation of the evolution of rotating stars.

The numerical construction of rotational equilibria has a long history.
There are actually many two-dimensional (2D) calculations under axisymmetry, all of them Eulerian, so far: for example,
\citet{Ostriker1968,Eriguchi1985,Hachisu1986} for the barotropic case, in which the pressure is a function of the density alone,
\citet{Jackson1970,Papaloizou1973,Eriguchi1991,Jackson2005} for the pseudo-barotropic case, where the \ac{EOS} is not barotropic but the isobaric and isopycnic surfaces coincide with each other in the star, and
\citet{Uryu1994,Uryu1995,Roxburgh2006,Fujisawa2015} for the baroclinic case, in which these surfaces are not aligned with each other in general; \citet{Espinosa2007,Espinosa2013} further considered meridional circulations.
Most of these studies employed an analytic first integral of the Euler equation, which is available only for slowly-rotating stars in perturbative methods \citep[e.g.,][]{Sharp1977} or for the (pseudo-)barotropic case.
It is emphasized again that the above works are all based on the Eulerian formulation and hence will not be suited for the evolutionary calculation.

Previously we developed a Lagrangian formulation on a triangular mesh and constructed some rotational equilibrium structures for both barotropic and baroclinic \ac{EOS}'s \citep[][]{Yasutake2015,Yasutake2016}. In these studies, the Lagrangian variational principle was adopted and structures with the minimal energy for given distributions of the mass, specific angular momentum and entropy for fluid elements were searched for. Although it worked in principle, it turns out that it is difficult to improve accuracy.

In this study, we develop a new Lagrangian formulation, in which the force-balance equations are solved to obtain the positions of fluid elements with given triples of the mass, specific angular momentum and entropy, that give a rotational equilibrium as a whole. Augmented with a remapping scheme, which avoids a mesh distortion, the new method is more accurate than the previous one. The purpose of this paper is to give a detailed description of this new formulation and demonstrate its performance: we construct both barotropic and baroclinic rotational equilibria with successively increasing angular momenta for three \ac{EOS}'s with different stiffnesses; we also study the resolution dependence; we finally apply the method to a toy-model calculation of the cooling of a rotational \ac{WD}. The incorporation of more detailed physics will be a future task.

This paper is organized as follows. In section~\ref{sec:method}, we describe the formulation in detail. The numerical models are explained in section~\ref{sec:model}, and their results are presented in section~\ref{sec:result}. Finally section~\ref{sec:conclusion} is a summary of this paper.

\section{Methods}
\label{sec:method}

In this section, we give the basic equations and describe how to solve them for the positions of fluid elements, or those of the Lagrangian grid points in the finite-differenced version. Throughout this paper we assume axisymmetry and equatorial symmetry; we ignore possible meridian fluid motions such as convection and circulation, assuming permanent rotation; magnetic fields are also neglected. These issues will be addressed in the subsequent papers.

\subsection{Force-balance equations}

Here we give the force-balance equations on the 2D Lagrangian coordinates deployed in the meridian section.
In contrast to the Eulerian coordinates, which are fixed to space, the Lagrangian coordinates are attached to fluid. As the fluid moves, so do the coordinates. In the stellar evolution, the star contracts or expands very slowly as it evolves. In the Lagrangian formulation, the coordinates also shrink or spread so that the coordinates of each fluid element should be unchanged.

In our formulation, we first consider a spherical reference configuration, which serves as the Lagrangian coordinates, and set the profiles of mass, specific entropy and specific angular momentum on it and fix them; we then seek a (generally non-spherical) configuration in mechanical equilibrium for these profiles on the fluid elements so that the correspondence of the fluid element and these physical quantities should be unchanged in the new configuration.

This may be more easily understood in the finite-differenced version. As shown in Fig.~1, the reference configuration is discretized on the (ordinary) spherical coordinates $(r, \theta)$ in the meridian section (shown with blue dashed lines) and the grid points are regarded as the finite-differenced version of fluid elements; to each of them we assign a mass, specific entropy and specific angular momentum; we then try to find their spatial positions $(R, \Theta)$ in the equilibrium configuration (depicted with red dashed lines), which is normally non-spherical, with those physical parameters allotted to them unchanged. In the following we explain how this is done.

We use $(r, \theta)$ on the reference configuration as the Lagrangian coordinates to label the fluid elements. Eventually we write the basic equations with these Lagrangian coordinates but we begin with the familiar force balance equations on the Euler coordinates, $(R, \Theta)$
\footnote{This is a bit abuse of notation. In fact, $(R, \Theta)$ are the values of the Euler coordinates that the fluid elements labeled with the Lagrangian coordinates $(r, \theta)$ have in the rotational equilibrium. See the explanation given below.}:
\begin{equation}
    \frac{1}{\rho}\bm\nabla P = -\bm\nabla\phi + \frac{1}{2}\Omega^2\bm\nabla(R\sin\Theta)^2,
    \label{eq:Euler}
\end{equation}
where $\bm\nabla$ is the operator of the derivative. They are decomposed into the radial and angular components,
\begin{align}
	F_R &\equiv \frac{\partial P}{\partial R} + \rho \frac{\partial \phi}{\partial R} - \rho (R\sin\Theta) \Omega^2 \sin\Theta = 0,
	\label{eq:Euler_R}  \\
	F_{\Theta} &\equiv \frac{\partial P}{R \partial \Theta} + \rho \frac{\partial \phi}{R \partial \Theta} - \rho (R\sin\Theta) \Omega^2 \cos\Theta = 0,%  \\
	\label{eq:Euler_T}
\end{align}
where $\rho, P, \phi, \Omega$ are the density, the pressure, the gravitational potential and the angular velocity, respectively.
We rewrite these equations in terms of the Lagrangian coordinates $(r, \theta)$ by regarding the correspondence between $(r,\theta)$ and $(R, \Theta)$ as a coordinate transformation:
\begin{align}
    R &= R(r,\theta),   \\
    \Theta &= \Theta(r,\theta).
\end{align}

It is more convenient to use the following linear combinations of $F_R$ and $F_{\Theta}$ instead of themselves:
\begin{align}
    \left(
	\begin{array}{c}
		F_r	\\
		F_{\theta}
	\end{array}
    \right) = \left(
	\begin{array}{cc}
		\displaystyle \left. \frac{\partial R}{\partial r} \right. & \displaystyle \left. \frac{R\partial \Theta}{\partial r} \right. 	\\
		\\
		\displaystyle \left. \frac{\partial R}{r\partial \theta} \right. & \displaystyle \left. \frac{R\partial \Theta}{r\partial \theta} \right.
	\end{array}
	\right)\left(
	\begin{array}{c}
		F_R	\\
		F_{\Theta}
	\end{array}
    \right).
    \label{eq:jac_eq}
\end{align}
They are written explicitly as follows:
\begin{align}
	F_{r}
	&= \left[ \left( \frac{\partial R}{\partial r} \right)\frac{\partial P}{\partial R}+\left( \frac{R\partial \Theta}{\partial r} \right)\frac{\partial P}{R\partial \Theta} \right]  \notag \\
	&\hspace{1cm} +\rho \left[\left( \frac{\partial R}{\partial r} \right)\frac{\partial \phi}{\partial R}+\left( \frac{R\partial \Theta}{\partial r} \right)\frac{\partial \phi}{R\partial \Theta} \right]   \notag \\
	&\hspace{1cm} + \rho(R\sin\Theta)\Omega^2 \left[ \left( \frac{\partial R}{\partial r} \right)\sin\Theta+\left( \frac{R\partial \Theta}{\partial r} \right)\cos\Theta \right] \notag	\\
	&= \frac{dP}{dr}
	+\rho \left[\left( \frac{\partial R}{\partial r} \right)F_{\mathrm{grav}}^{(R)}+\left( \frac{R\partial \Theta}{\partial r} \right)F_{\mathrm{grav}}^{(\Theta)} \right] \notag \\
	&\hspace{1cm} + \frac{\rho j^2}{(R\sin\Theta)^3} \left[ \left( \frac{\partial R}{\partial r} \right)\sin\Theta+\left( \frac{R\partial \Theta}{\partial r} \right)\cos\Theta \right] = 0,
	\label{eq:differenced_r}
\end{align}
\begin{align}
	F_{\theta}
	&= \left[ \left( \frac{\partial R}{r\partial \theta} \right)\frac{\partial P}{\partial R}+\left( \frac{R\partial \Theta}{r\partial \theta} \right)\frac{\partial P}{R\partial \Theta} \right]    \notag \\
	&\hspace{1cm} + \rho \left[ \left( \frac{\partial R}{r\partial \theta} \right)\frac{\partial \phi}{\partial R}+\left( \frac{R\partial \Theta}{r\partial \theta} \right)\frac{\partial \phi}{R\partial \Theta} \right] \notag \\
	&\hspace{1cm} + \rho(R\sin\Theta)\Omega^2 \left[ \left(\frac{\partial R}{r\partial \theta} \right)\sin\Theta+\left( \frac{R\partial \Theta}{r\partial \theta} \right)\cos\Theta \right] \notag	\\
	&= \frac{dP}{rd\theta}
	+ \rho \left[ \left( \frac{\partial R}{r\partial \theta} \right)F_{\mathrm{grav}}^{(R)}+\left( \frac{R\partial \Theta}{r\partial \theta} \right)F_{\mathrm{grav}}^{(\Theta)} \right]   \notag \\
	&\hspace{1cm} + \frac{\rho j^2}{(R\sin\Theta)^3} \left[ \left(\frac{\partial R}{r\partial \theta} \right)\sin\Theta+\left( \frac{R\partial \Theta}{r\partial \theta} \right)\cos\Theta \right] = 0,
	\label{eq:differenced_t}
\end{align}
where $j$ is the specific angular momentum, and $F_{\mathrm{grav}}^{(R)}$ and $F_{\mathrm{grav}}^{(\Theta)}$ are the $R$- and $\Theta$-components of the gravitational force, respectively (see section~\ref{sec:gravity} for details).
In this formulation, we will solve these equations to obtain the functional forms of $R(r, \theta)$ and $\Theta (r, \theta)$. This is actually done in the finite-difference approximation, i.e., we will seek for the values of $R$ and $\Theta$ for discrete grid points on the Lagrangian coordinates $(r, \theta)$.
\begin{figure}
	\centering
	\includegraphics[width=0.8\columnwidth]{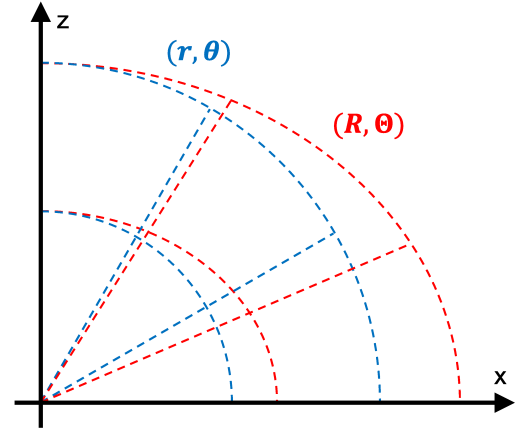}
	\caption{Schematic picture of the Lagrangian coordinates $(r, \theta)$ defined for the reference configuration and their actual configuration $(R, \Theta)$ in rotational equilibrium.}
	\label{fig:schematic}
\end{figure}

\begin{figure*}
	\centering
	\includegraphics[keepaspectratio,scale=0.35]{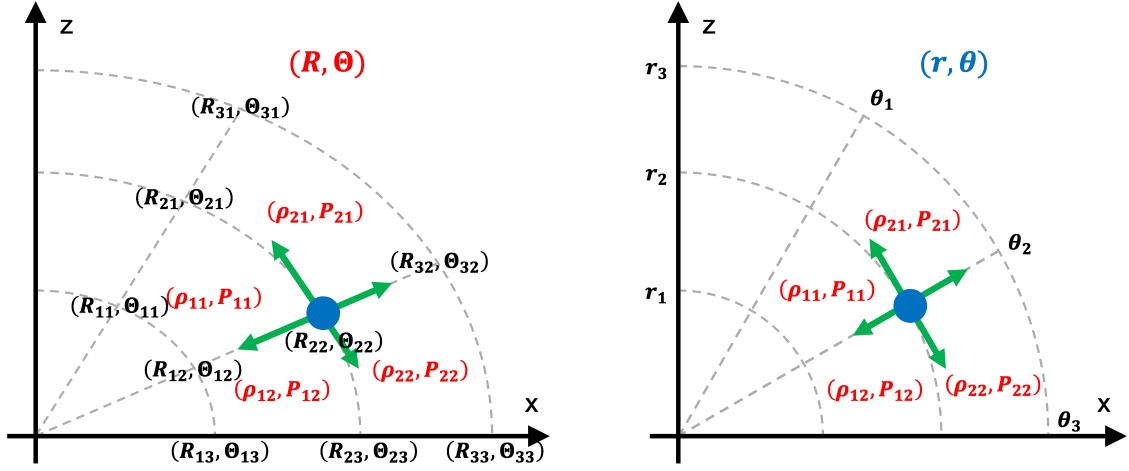}
	\caption{Assignments of various quantities in the real (left panel) and reference (right panel) configurations. See equations (\ref{eq:dif_r}) and (\ref{eq:dif_t}).
	}
	\label{fig:differentiation}
\end{figure*}

The finite-differenced force-balance equations at the blue dot in Fig.~\ref{fig:differentiation} are given as
\begin{align}
	F_r
	&= \frac{P_{21}+P_{22}-P_{11}-P_{12}}{r_3-r_1}  \notag \\
	&\hspace{0.5cm} + \rho\left[\left( \frac{R_{32}-R_{12}}{r_3-r_1} \right) F_{\mathrm{grav}}^{(R)} + \left( \frac{R_{22}(\Theta_{32}-\Theta_{12})}{r_3-r_1} \right) F_{\mathrm{grav}}^{(\Theta)}\right]	\notag	\\
	&\hspace{0.5cm} +\frac{\rho j^2}{(R_{22}\sin\Theta_{22})^3} \left[ \left( \frac{R_{32}-R_{12}}{r_3-r_1} \right)\sin\Theta_{22} \right.    \notag \\
	&\hspace{2cm} \left. +R_{22} \left( \frac{R_{22}(\Theta_{32}-\Theta_{12})}{r_3-r_1} \right)\cos\Theta_{22} \right] = 0,
	\label{eq:dif_r}
\end{align}
\begin{align}
	F_{\theta}
	&= \frac{P_{12}+P_{22}-P_{11}-P_{21}}{r_2(\theta_3 - \theta_1)} \notag \\
	&\hspace{0.5cm} +\rho\left[\left( \frac{R_{23}-R_{21}}{r_2(\theta_3 - \theta_1)} \right)F_{\mathrm{grav}}^{(R)} + \left( \frac{R_{22}(\Theta_{23}-\Theta_{21})}{r_2(\theta_3-\theta_1)} \right) F_{\mathrm{grav}}^{(\Theta)}\right] \notag	\\
	&\hspace{0.5cm} +\frac{\rho j^2}{(R_{22}\sin\Theta_{22})^3} \left[ \left( \frac{R_{23}-R_{21}}{r_2(\theta_3-\theta_1)} \right)\sin\Theta_{22} \right. \notag \\
	&\hspace{2.5cm} \left. +\left( \frac{R_{22}(\Theta_{23}-\Theta_{21})}{r_2(\theta_3-\theta_1)} \right) \cos\Theta_{22} \right] = 0.
	\label{eq:dif_t}
\end{align}
The notations adopted in these equations are also given in Fig.~\ref{fig:differentiation};
$F_{\mathrm{grav}}^{(R)}, F_{\mathrm{grav}}^{(\Theta)}$ are evaluated at the blue circle (see section~\ref{sec:gravity} for details) and so is $j$; $\rho=(\rho_{11}+\rho_{12}+\rho_{21}+\rho_{22})/4$.
Note that the force-balance equations are evaluated at the grid points; the density and pressure are defined at the cell centers whereas the specific entropy, the angular velocity (and hence the specific angular momentum as well) and the gravitational forces are given on the grid points.

The density attached to the cell is obtained by dividing the mass $\Delta m$ assigned to the cell by the cell volume $\Delta V$, which is given by
\begin{equation}
    \Delta V = |\mathrm{det}\ {\bf J}|\ \Delta v,
\end{equation}
where ${\bf J}$ is the Jacobian \footnote{Strictly speaking, this is the ratio of the infinitesimal volumes, i.e., the Jacobian with the measures included.} for the coordinate transformation $(r, \theta) \rightarrow (R, \Theta)$ given as
\begin{align}
	{\bf J} = \left[
	\begin{array}{cc}
		\displaystyle \left. \frac{\partial R^3/3}{\partial r^3/3} \right|_{\cos\theta} & \displaystyle \left. \frac{\partial \cos\Theta}{\partial r^3/3} \right|_{\cos\theta}	\\
		\displaystyle \left. \frac{\partial R^3/3}{\partial \cos\theta} \right|_{r^3/3} & \displaystyle \left. \frac{\partial \cos\Theta}{\partial \cos\theta} \right|_{r^3/3}
	\end{array}
	\right],
\end{align}
and $\Delta v$ represents the volume of the corresponding cell in the reference configuration. Since the specific entropy is already given (on the grid points), we can obtain the pressure from the \ac{EOS}, which is assumed to be a function of the density and specific entropy. It is now apparent that once the positions of the Lagrangian grid points, i.e., $(R, \Theta)$ are given as a function of $(r, \theta)$, all the thermodynamic quantities are obtained. They are hence functionals of the coordinate-transformation functions, $R(r, \theta)$ and $\Theta(r, \theta)$. It is also self-evident that the gravitational and centrifugal forces are also functionals of the same transformation functions. If this transformation is correctly given, the quantities so obtained satisfy the force-balance equations, equations (\ref{eq:differenced_r}) and (\ref{eq:differenced_t}) (or equations (\ref{eq:dif_r}) and (\ref{eq:dif_t}) for the finite-differeced version). Our task is hence to find such coordinate transformations.

In the actual numerical computations all physical quantities are converted to dimensionless ones with the stellar radius $r_s$ in the reference configuration, the stellar mass $M_{\mathrm{tot}}$, and the gravitational constant $G$ as follows:
\begin{align}
    \hat{r} &\equiv \frac{r}{r_s},    \\
    \hat{m} &\equiv \frac{m}{M_{\mathrm{tot}}}, \\
    \hat{\rho} &\equiv \frac{\rho}{M_{\mathrm{tot}}/r_s^3},  \\
    \hat{P} &\equiv \frac{P}{GM_{\mathrm{tot}}^2/r_s^4},   \\
    \hat{K_0} &\equiv \frac{K_0}{GM_{\mathrm{tot}}^{1-1/N}r_s^{-1+3/N}},  \label{eq:normK} \\
    \hat{\Omega} &\equiv \frac{\Omega}{\sqrt{GM_{\mathrm{tot}}/r_s^3}},    \\
    \hat{\phi} &\equiv \frac{\phi}{GM_{\mathrm{tot}}/r_s},  \\
    \hat{j} &\equiv \frac{j}{\sqrt{r_s GM_{\mathrm{tot}}}}, \\
    \hat{J_{\mathrm{tot}}} &\equiv \frac{J_{\mathrm{tot}}}{\sqrt{r_s GM_{\mathrm{tot}}^3}},
\end{align}
where hat denotes dimensionless quantities; $K_0$ and $N$ are the constants in the polytropic \ac{EOS} $P=K_0\rho^{1+1/N}$; $j$ and $J_{\mathrm{tot}}$ are the specific angular momentum and the total angular momentum (see section~\ref{sec:model} for more details).

\subsection{Self-gravity}
\label{sec:gravity}

In this subsection we describe the calculation of self-gravity more in detail. It turns out that the precise computation of the gravitational potential and force is crucially important in obtaining rotational equilibria with our method. We employ the spectral method based on the Legendre functions.
We solve the Poisson equation for the gravitational potential on the Eulerian coordinates. This is because the Laplacian takes the  simplest form and the application of the spectral method becomes easiest. Note that we alternately solve the force-balance equations and the Poisson equation repeatedly until a convergence is obtained (see section~\ref{sec:numerical_method} for more). Since the Lagrangian grid points change their positions on the Eulerian coordinates, the gravitational potentials at their positions vary also although the potential is unchanged as a function on the Eulerian coordinates. Since the spectral method gives not the values of the potential on the grid points but the function itself, the evaluation of the potentials at the positions of the Lagrangian grid points is straightforward and accurate.

The Poisson equation is written as
\begin{equation}
	\frac{\partial}{\partial R}\left( R^2 \frac{\partial\phi}{\partial R} \right) + \frac{\partial}{\partial\mu}\left[(1-\mu^2)\frac{\partial\phi}{\partial\mu}\right] = 4 \pi G \rho(R,\mu) R^2,
	\label{eq:Poisson}
\end{equation}
where $\mu$ is the cosine of the zenith angle, $\mu=\cos\Theta$.
As mentioned above, we expand the gravitational potential (and the density) as follows:
\begin{align}
	\phi(R,\mu) = \sum_{l=0}^{l_{\mathrm{max}}}\sum_{m=0}^{m_{\mathrm{max}}} a_{lm} \tilde{P}_l(R) P_m(\mu),
	\label{eq:Lege_phi} \\
	4\pi G\rho(R,\mu)R^2 = \sum_{l=0}^{l_{\mathrm{max}}}\sum_{m=0}^{m_{\mathrm{max}}} b_{lm} \tilde{P}_l(R) P_m(\mu)
    \label{eq:Lege_rho}
\end{align}
where $l_{\mathrm{max}}=22, m_{\mathrm{max}}=24$ are employed for all our models and $P_m$ are the ordinary Legendre polynomials defined on the domain $[-1, 1]$ whereas $\tilde{P}_l$ are their shifted variants with the domain $[0, R_*]$, in which $R_*$ is the stellar radius at the equator; they are defined as $\tilde{P}_l(R) = P_l(x)$ with $R=R_*(x+1)/2$ for $x\in[-1,1]$. Then the orthogonality relations are modified as
\begin{equation}
    \int_0^{R_*}\tilde{P}_l(R)\tilde{P}_m(R)dR = \frac{R_*}{2l+1}\delta_{lm}.
    \label{eq:Dlm}
\end{equation}
Substituting equations (\ref{eq:Lege_phi}), (\ref{eq:Lege_rho}) into equation (\ref{eq:Poisson}) and using equation (\ref{eq:Dlm}), one obtains
\begin{align}
    \sum_{l=0}^{l_{\mathrm{max}}}\left\{ \frac{\partial}{\partial R}\left( R^2 \frac{\partial \tilde{P}_l(R)}{\partial R} \right) - m(m+1)\tilde{P}_l(R) \right\}a_{lm} = \sum_{l=0}^{l_{\mathrm{max}}}b_{lm}\tilde{P}_l(R).
    \label{eq:expansion}
\end{align}

The radial derivative and the multiplication of $R^2$ in the above equation are further expanded as
\begin{align}
    \frac{\partial \tilde{P}_{l}(R)}{\partial R} = \sum_{m=0}^{l-1}C_{lm}\tilde{P}_m(R), \\
	R^2 \tilde{P}_l(R) = \sum_{m=0}^{l+2}D_{lm}\tilde{P}_m(R),
	\label{eq:r2}
\end{align}
where $C_{lm}$ and $D_{lm}$ are numerical constants; $C_{lm}$ is given as
\begin{align}
  C_{lm} = \left[
    \begin{array}{cccccc}
        0      & 1       & 0       & 1      & 0 & \cdots \\
        \vdots & 0       & 3       & 0      & 3 & \cdots \\
               & \vdots  & 0       & 5      & 0 & \cdots \\
	           &         & \vdots  & 0      & 7 & \cdots \\
		       &         &         & \vdots & 0 & \cdots \\
		       &         &         &        &   & \ddots
    \end{array}
  \right]
\end{align}
from the following relation:
\begin{equation}
    \frac{d}{dx}P_{n+1}(x)=(2n+1)P_n(x)+(2n-3)P_{n-2}(x)+(2n-7)P_{n-4}(x)+\cdots;
\end{equation}
$D_{lm}$ is obtained as
\begin{align}
    D_{lm} = R_*^2\left[
    \begin{array}{ccccccc}
        1/3    & 1/6    & 1/30   & 0      & \cdots &        &        \\
        1/2    & 2/5    & 1/5    & 3/70   & 0      & \cdots &        \\
        1/6    & 1/3    & 8/21   & 3/14   & 1/21   &  0     & \cdots \\
	    0      & 1/10   & 3/10   & 17/45  & \ddots & \ddots &        \\
		\vdots & 0      & \ddots & \ddots & \ddots &        &        \\
		       & \vdots &        &        &        &        &
    \end{array}
  \right],
\end{align}
from the following integral:
\begin{equation}
	D_{lk} = \frac{2k+1}{R_*} \int_0^{R_*} R^2 \tilde{P}_l(R) \tilde{P}_k(R) dR;
\end{equation}
their concrete expressions are given, for example, in \citet{arfken1985} (P.700) as
\begin{align}
    D_{l\ l-2} &= \frac{(2l+1)R_*^2}{8}\frac{2l(l-1)}{(2l-3)(2l-1)(2l+1)},   \\
    D_{l\ l-1} &= \frac{(2l+1)R_*^2}{8}\frac{4l}{(2l-1)(2l+1)},   \\
    D_{l\ l  } &= \frac{(2l+1)R_*^2}{8}\left[\frac{2}{2l+1}+\frac{2(2l^2+2l-1)}{(2l-1)(2l+1)(2l+3)}\right],   \\
    D_{l\ l+1} &= \frac{(2l+1)R_*^2}{8}\frac{4(l+1)}{(2l+1)(2l+3)},   \\
    D_{l\ l+2} &= \frac{(2l+1)R_*^2}{8}\frac{2(l+1)(l+2)}{(2l+1)(2l+3)(2l+5)}.
\end{align}

The first term on the left-hand side of equation (\ref{eq:expansion}) can be rewritten as follows:
\begin{align}
	&\sum_{l=0}^{l_{\mathrm{max}}} \frac{\partial}{\partial R} \left( R^2 \frac{\partial \tilde{P}_{l}(R)}{\partial R} \right) a_{lm}    \notag \\
	&\hspace{1cm} = \sum_{l=0}^{l_{\mathrm{max}}} \sum_{n=0}^{l_{\mathrm{max}}-1} \sum_{i=0}^{l_{\mathrm{max}}+1} \sum_{j=0}^{l_{\mathrm{max}}} \left( a_{lm} C_{ln} D_{ni} C_{ij} \tilde{P}_j(R) \right).
\end{align}
Using again the orthogonality relations, (\ref{eq:Dlm}), we finally obtain the following equation for the expansion coefficients of $a_{lm}$:
\begin{equation}
	\sum_{l=0}^{l_{\mathrm{max}}} a_{lm'} \left\{ \sum_{n=0}^{l_{\mathrm{max}}-1} \sum_{i=0}^{l_{\mathrm{max}}+1} C_{ln} D_{ni} C_{ij} - m'(m'+1)\delta_{lj} \right\} = b_{m'j}.
    \label{eq:Mjm}
\end{equation}
We define $M_{lj}=\sum_n \sum_i C_{ln}D_{ni}C_{ij} - m'(m'+1)\delta_{lj}$.

The evaluation of $b_{lm}$ is a bit tricky, since the density distribution is given on the Lagrangian coordinates. The integration for $b_{lm}$
\begin{align}
    b_{lm} &= \frac{2l+1}{R_*}\frac{2m+1}{2} \int_0^{R_*} \int_{-1}^1 4\pi Gr^2\rho(R,\mu) \tilde{P}_l(R)P_m(\mu)dRd\mu
\end{align}
is conducted with the Gaussian quadrature method. In so doing, we utilize the interpolation scheme developed for remapping (see section~\ref{sec:redis} for details) to obtain the values of density at the quadrature points.

Equation (\ref{eq:Mjm}) now written as $\sum_l a_{lm'}M_{lj}=b_{m'j}$ can be solved by inverting the matrix $M_{li}$ with a suitable method. Before doing so, however, we need to take a proper account of the inner and outer boundary conditions, which are given for $\phi_m=\sum_l a_{lm}\tilde{P}_l(R)$ as
\begin{align}
    &\left.\frac{\partial\phi_m}{\partial R}\right|_{R=0} = 0,
    \label{eq:BC1}  \\
    &\Rightarrow \sum_l a_{lm} \times (-1)^l l(l+1) = 0,
    \label{eq:BC1_mat}
\end{align}
and
\begin{align}
    &\left.\frac{\partial\phi_m}{\partial R}\right|_{R=R_*} = -\frac{m+1}{R_*}\phi_m,
    \label{eq:BC2}  \\
    &\Rightarrow \sum_l a_{lm} \times \{(m+1)+(l-1)l\} = 0,
    \label{eq:BC2_mat}
\end{align}
for $0\leq m\leq m_{\mathrm{max}}$. In fact, the original matrix $M_{lj}$ is singular as it is. We need to replace its two columns with the above boundary conditions, (\ref{eq:BC1_mat}) and (\ref{eq:BC2_mat}). The $m$-th column should be replaced with equation (\ref{eq:BC2_mat}); although here is some arbitrariness in the choice of the other column, we found it best to replace the last column with equation (\ref{eq:BC1_mat}). The resultant matrix is non-singular and can be inverted without difficulties.

The $R$- and $\Theta$-components of the gravitational force are obtained by further differentiating the potential thus obtained:
\begin{align}
	F_{\mathrm{grav}}^{(R)}(R,\mu) &= \frac{\partial\phi}{\partial R}
	= \sum_{l=0}^{l_{\mathrm{max}}-1} \sum_{m=0}^{m_{\mathrm{max}}} \sum_{k=0}^{l_{\mathrm{max}}} C_{lk} a_{km} \tilde{P}_l(R) P_m(\mu), \notag  \\
	&= \sum_{l=0}^{l_{\mathrm{max}}-1} \sum_{m=0}^{m_{\mathrm{max}}} G_{lm}^{(R)} \tilde{P}_l(R) P_m(\mu),    \\
	F_{\mathrm{grav}}^{(\Theta)}(R,\mu) &= \frac{\partial\phi}{R\partial\Theta} = \frac{\sin\Theta}{R}\frac{\partial\phi}{\partial\cos\Theta},    \notag \\
	&= \frac{\sin\Theta}{R} \sum_{l=0}^{l_{\mathrm{max}}} \sum_{m=0}^{m_{\mathrm{max}}-1} \sum_{k=0}^{m_{\mathrm{max}}} a_{lk} C_{km} \tilde{P}_l(R) P_m(\mu), \notag \\
	&= \frac{\sin\Theta}{R} \sum_{l=0}^{l_{\mathrm{max}}} \sum_{m=0}^{m_{\mathrm{max}}-1} G_{lm}^{(\Theta)} \tilde{P}_l(R) P_m(\mu).
\end{align}
The expansion coefficients $G^{(R)}_{lm}$ and $G^{(\Theta)}_{lm}$ in the above equations are given as
\begin{align}
    G^{(R)}_{lm} &= \sum_{k=0}^{l_{\mathrm{max}}} a_{km}C_{kl},    \\
    G^{(\Theta)}_{lm} &= \sum_{k=0}^{m_{\mathrm{max}}} a_{lk}C_{km}.
\end{align}

One last care is needed. As we mentioned earlier, the force-balance equations and the Poisson equation are solved alternately. Since the gravitational potential is fixed in solving the force-balance equations\footnote{The gravitational potential is fixed as a function but its values at the Lagrangian grid points are changed according to their motions.}, it happens that some Lagrangian grid points near the stellar surface get out of the outer boundary for the potential calculation. We then need to extend the potential outward. This should be done so that the potential be connected continuously with the solution in vacuum:
\begin{align}
    \phi(R,\mu) = -\sum_{m=0}^{m_{\mathrm{max}}} \frac{GM_m}{R^{m+1}} P_m(\mu).
    \label{eq:BC_out}
\end{align}
The value of $M_m$ is set so that the continuity should be guaranteed. In fact, we determine $M_m$ from equations (\ref{eq:Lege_phi}) and (\ref{eq:BC_out}) as
\begin{equation}
    M_m = \frac{R_*^{m+1}}{G}\times\sum_{l=0}^{l_{\mathrm{max}}} a_{lm} \tilde{P}_l(R_*).
\end{equation}

\subsection{Remapping}
\label{sec:redis}

\begin{figure}
	\centering
	\includegraphics[width=\columnwidth]{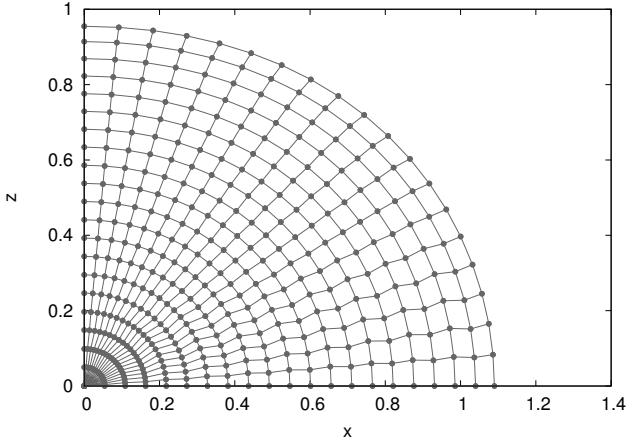}
	\caption{A grid configuration after the equlibrium is reached without remapping for the model with the polytropic index $N=1.5$ and $(N_r, N_{\theta})=(20,20)$.}
	\label{fig:distortN1.5}
\end{figure}

One of the difficulties with the Lagrange formulation is the deformation of the mesh. As mentioned earlier, the Lagrangian grid moves with the matter and we do not know a priori what the final mesh configuration should be like. We observed indeed that, starting from a spherical configuration of reference, the mesh becomes flattened as it approaches the rotational equilibrium for given profiles of the mass, specific entropy and specific angular momentum on the Lagrangian grid. We also found that the grid tends to become zigzag as it gets nonspherical (see Fig.~\ref{fig:distortN1.5}). This is due to the numerical errors in finite-differencing caused by the grid deformation. It leads not only to inaccuracies of the derived configuration but also to non-convergence of the iterative procedure eventually. We hence need to implement a regridding to repair the grid deformation accompanied by a redistribution of the conserved quantities, i.e., the mass, specific entropy and specific angular momentum on the new grid, the procedure we refer to as the remapping.

The remapping hence consists of the two processes: (1) the generation of a new smooth grid and (2) the redistribution of the conserved quantities on the new grid. This is actually performed via a couple of steps as follows:
\begin{enumerate}
    \item The construction of the new grid begins with the surface-fitting with a polynomial. We put new grid points on this surface uniformally (normally) and connect them with the origin to define the new straight radial rays. We then deploy new radial grid points on these rays so that they come close to the original points. We thus obtain the new mesh $(r', \theta')$.

    \item We calculate the masses to be assigned to cells of the new mesh.
    This is done simply as an interpolation of density to the new cell center. In so doing we assume that the density is uniform in each old cell.
    The density so obtained at the new cell center is then multiplied by the new cell volume to obtain its mass.
    We sum up these masses to obtain the total mass, which is normally not equal to the original value completely. We then multiply a correcting factor (common to all cells) to guarantee the mass conservation.

    \item The specific entropy is interpolated in the same way but to the new grid points.

    \item We then proceed to the redistribution of the specific angular momentum. We find that it is better to use the angular velocity instead of the specific angular momentum for interpolation. This is because the angular momentum decreases rapidly to 0 on the rotation axis, giving large interpolation errors.
    The angular velocities interpolated to the new grid points are converted to the specific angular momentum there.
    The angular velocity on a new grid point is interpolated from the nearby nine old grid points denoted by $(r_{jk},\theta_{jk})$ as
    \begin{equation}
        \Omega'(r',\theta') = \sum_{j=1}^3 \sum_{k=1}^3 \hat{M}_j(\alpha) \hat{M}_k(\beta) \Omega(r_{jk},\theta_{jk}),
    \end{equation}
    where $\hat{M}_j$ $(j=1,3$) are the so-called shape functions with $\alpha, \beta$ being the natural coordinates $-1\leq\alpha, \beta\leq 1$. For details we refer to our associated paper \citep{Okawa2022}.
    We sum them up after multiplying them with the corresponding densities to obtain the total angular momentum. Again, since it is slightly different from the original value normally, we multiply a correcting factor to guarantee the conservation of the total angular momentum.
\end{enumerate}

The remapping is actually a part of the convergence calculations, in which the force-balance equations and the Poisson equation for self-gravity are solved alternately till the convergence is obtained. It is administered every time this convergence is reached (see also section~\ref{sec:numerical_method} below).

\subsection{Numerical method}
\label{sec:numerical_method}

We solve the finite-differenced force-balance equations (\ref{eq:dif_r}) and (\ref{eq:dif_t}) for $R_{ij}$ and $\Theta_{ij}$ iteratively, using the W4 method \citep[][]{Okawa2023}, which we have developed as a new root finder of nonlinear simultaneous equations that could replace the Newton-Raphson method when it fails. In fact, we found it all but impossible to obtain a convergence of the Newton-Raphson iterations for our equations, which are very stiff indeed \citep[see][]{Okawa2022}.
The W4 method has a local convergence property similar to that of the Newton-Raphson method but is superior in the global convergence \citep[][]{Okawa2023}. It has several variants and we use in this paper the one called W4LH, which utilizes the LH decomposition of the Jacobian matrix (see appendix in \citet{Fujisawa2019} for more details).

The actual numerical procedure to obtain a rotational equilibrium for a given profile of the mass, specific entropy and specific angular momentum on the Lagrangian grid is based on the self-consistent-field scheme \citep[][]{Ostriker1968}, i.e., the self-gravity and the matter positions are solved iteratively until they are no longer changed.
The gravitational potential, and hence the gravitational forces as well, are derived for a temporal matter profile in the method described in section~\ref{sec:gravity}. The gravitational potential is then fixed when the force-balance equations are solved. Note that since the Lagrangian grid points move around, the gravitational forces at their positions are changed in this calculation and are updated according to the position while the gravitational potential remains fixed. Once the new Lagrangian grid is obtained, we calculate the new gravitational potential for the matter profile derived from this new configuration.
We repeat this procedure until the gravitational potential and the grid positions are essentially unchanged in each step any longer. At this point we perform the remapping to correct a (normally small) deformation of the grid. We then return to the alternate solutions of the force-balance equations and the Poisson equation for self-gravity. The remapping is administered again after the convergence. This outer loop is also iterated until the remapping no longer changes the grid configuration. We find convergence indeed both in the inner and outer loops if the initial configuration is close to the equilibrium. We normally guarantee this condition by changing the conserved quantities gradually from one model to another and using the equilibrium configuration in the previous model as the initial configuration for the next model.

In order to validate this new formulation, we check the accuracy of solutions. As one of the diagnostics, we use the Virial relation \citep[see appendix \ref{app:Virial};][]{Eriguchi1985,Fujisawa2015,Yasutake2015,Yasutake2016}.
We define the Virial constant $V_c(\geq 0)$ as follows:
\begin{equation}
    V_c = \frac{\left|3U  + W + 2T\right|}{\left|U\right|+\left|W\right|+\left|T\right|},
    \label{eq:Virial}
\end{equation}
where $U, W$ and $T$ are the integrated pressure, the gravitational energy and the rotational energy, respectively (see appendix~\ref{app:Virial} for the definitions). Since this quantity $V_c$ vanishes for the exact solution, the smaller it is, the more accurate the numerical solution is.
The integrations are numerically conducted on the same Lagrangian mesh employed to derive the rotational equilibrium. We also check the well-known fact that the the rotation should be cylindrical, i.e., the angular velocity, and hence the specific angular momentum as well, are functions of the distance from the rotation axis alone for the (pseudo-)barotropic case (see appendix \ref{app:barotrope}). In the Lagrangian formulation, this is highly non-trivial, since the profile of the specific angular momentum in space (not on the Lagrangian grid) is determined by the final configuration of the Lagrangian grid points after the rotational equilibrium is established. In the baroclinic case, the rotation is non-cylindrical and obeys the so-called Bjerknes–Rosseland rule (see appendix \ref{app:barotrope}). We will confirm that it is satisfied in our baroclinic results. Finally, we construct with a well-calibrated Eulerian code \citep[][]{Fujisawa2015} the rotational equilibria that have the same mass and angular velocity distributions in space as the configurations derived with our Lagrangian code and compare them quantitatively.

\section{Models}
\label{sec:model}

In the following we describe the model calculations we performed in this paper to validate the new formulation and to demonstrate its capability in possible applications.

\subsection{Rotational Equilibria for Polyrtopes}
\label{sec:model_poly}

We begin with the barotropic case. We construct a sequence of rotating configurations with successively greater angular momenta. We employ the polytropic \ac{EOS}'s for simplicity as a representative of the barotropic \ac{EOS}:
\begin{equation}
	P = K_0 \rho^{1+\frac{1}{N}},
	\label{eq:polytrope}
\end{equation}
where $N$ is the polytropic index and $K_0$ is a constant. The latter may be interpreted as a (function of) specific entropy $s$ in an isentropic star. In fact, $K_0\propto \exp(s)$ in the case of the ideal gas.
In our formulation the specific entropy is one of the three quantities assigned to each fluid element (or to each Lagrangian grid point in the discretized version) and carried with it as it moves to its equilibrium position. This is a trivial issue, though, in the isentropic case considered here. In our model calculations, we consider three values of the polytropic index: $N = 1.0, 1.5, 2.5$, having in mind applications to various stars.

The specific angular momentum is another quantity attached to the fluid element and, as such, it is most conveniently specified on the Lagrangian grid in the reference configuration. In all models, we assume the following rotation law initially:
\begin{equation}
    j(r,\theta) = \frac{j_0(r\sin\theta)^2}{1+(r\sin\theta)^2},
    \label{eq:J_law}
\end{equation}
where $j_0$ is a constant to specify the overall rotation strength. The functional form is the same as the so-called $j$-constant law employed by \citet{Eriguchi1985} in their numerical constructions of rotational equilibria in their Eulerian formulation. Note that in our Lagrangian formulation, the actual rotation profile in the star is not known a priori but is obtained only after the equilibrium  configuration is established. In principle, there is no guarantee that it has the same functional form. Since the \ac{EOS} is barotropic, however, we know a priori that it is a function of the distance from the rotation axis alone (see appendix \ref{app:barotrope}). We emphasize again that this is a highly non-trivial issue for our Lagrangian formulation and can be used as a diagnostic. In the test calculations, we construct a series of rotational equilibria that has successively larger angular momenta, increasing the value of specific angular momentum of every fluid element (or Lagrangian grid point in the discretized version) gradually by a common factor\footnote{Note that the profile of the specific angular momentum in the reference configuration is no longer the one given in equation~(\ref{eq:J_law}) after remapping, since it re-grids also the reference configuration.}.

\subsection{Baroclinic Rotational Equilibria}
\label{sec:baroclinic}

Next we consider the baroclinic case, in which the isopycnic surfaces do not coincide with the isobaric surfaces and the rotation is not cylindrical, i.e., the angular velocity depends not only on the distance from the rotational axis but also on the height from the equator. This happens when the pressure is not a function of density alone, which is normally the case.

In order to consider such situations, we extend the polytropic \ac{EOS} as follows \citep[cf.][]{Fujisawa2015,Yasutake2016}:
\begin{align}
    \begin{split}
        &P(r,\theta) = K(r,\theta)\rho^{1+\frac{1}{N}}, \\
	    &K(r,\theta) = K_0\left[1+\epsilon_1\left(\frac{r}{R_{\mathrm{eq}}}\right)^2\sin^2\theta+\epsilon_2\left(\frac{r}{R_{\mathrm{eq}}}\right)^2\cos^2\theta\right],
    \end{split}
	\label{eq:baroclinic}
\end{align}
where $R_{\mathrm{eq}}$ is the equatorial radius and $\epsilon_1, \epsilon_2$ are positive constants less than 1, and represent the degree of baroclinicity. The expression is reduced to the polytrope when $\epsilon_1=\epsilon_2=0$. Note that $K(r, \theta)$ is given as a function of the Lagrangian coordinates. This may be interpreted as an assignment of specific entropies to the fluid elements. It should be clear that the specific entropy increases quadratically in the distance from the rotational axis if $\epsilon_1$ is non-vanishing and so does in the height from the equator if $\epsilon_2$ is non-zero. Once allotted, they are fixed in the subsequent search of the equilibrium configuration. Note that the spatial profile of the specific entropy is changed in the process. The specific angular momentum profile on the Lagrangian grid is the same as the polytropic model.

As mentioned above, the isobaric surfaces are inclined against the isopycnic surfaces according to the Bjerknes-Rosseland rule. We will check whether it is satisfied in the numerical results. The contours of the angular momentum are not parallel to the rotation axis and their inclinations can be predicted in a similar way, which we will also employ as a diagnostic in this test calculation.

\subsection{Cooling of a Rotating White Dwarf}

This toy model calculation is meant to demonstrate the capability of our new formulation to calculate the cooling of rotating stars that occurs over the secular time scale. Here we pick up \acp{WD}.

The number of high-mass \acp{WD} observed is increasing rapidly these days \citep[][]{Gaia2018,Kilic2021}. Some of them are thought to have been generated as a result of the merger of two \acp{WD} and are expected to be rotating rapidly due to the conservation of angular momentum. We consider here the cooling of such a rapidly rotating \ac{WD}. In the toy model, we describe the evolution via cooling as a sequence of the rotational equilibria that have the same mass and specific angular momentum profiles on the Lagrangian coordinates but have successively smaller specific entropies in each fluid element due to cooling. Note that the angular momentum distribution in space does change in this sequence as the rotational configurations become more flattened. We ignore a possible angular momentum transport among fluid elements as well as convection and other meridian motions for simplicity.

The initial condition is constructed as follows. We employ the 1D stellar evolution code MESA \citep[][]{MESA2011,MESA2013,MESA2015,MESA2018,MESA2019} to calculate the evolution of the non-rotating star with the zero-age \ac{MS} mass of $6~\mathrm{M}_{\odot}$ with the solar metallicity up to the formation of a \ac{WD} of $\sim0.6~\mathrm{M}_{\odot}$ at about $3.48\times 10^{8}$ yrs.
Although MESA incorporates a tabulated \ac{EOS} that takes into account various physics, we adopt here a simplified \ac{EOS} given as
\begin{align}
    \begin{split}
        &P = K(m) \rho^{1+\frac{1}{N}},  %\\
    \end{split}
	\label{eq:cooling}
\end{align}
in which $N=2.5$ and the value of $K$ is obtained from the MESA result as $K(m)=P(m)/\rho(m)^{1+1/N}$ as a function of mass coordinates $m$. To save the computation time, we utilize the polytropic model with the same value of $N$ constructed in section~\ref{sec:model_poly}. We change $K$ gradually from the constant value in the polytropic model to the value given in equation~(\ref{eq:cooling}). In so doing, we calculate the mass coordinate of each fluid element in the (spherical) reference configuration. We thus obtain the rotational \ac{WD} at the initial time.

In this toy model, the temporal evolution of $K$ is set by hand as follows. We run MESA again to follow the cooling of the same (non-rotating) \ac{WD} for another $10^8$ yrs in 1D. Employing the initial and final values of $K$ of the individual fluid elements in this additional 1D simulation, we give the time evolution of $K$ at each grid point in the Lagrangian coordinates as
\begin{equation}
    K(m) = (1-\alpha) K_{\mathrm{ini}}(m) + \alpha K_{\mathrm{fin}}(m),
    \label{eq:dis_K}
\end{equation}
where $K_{\mathrm{ini}}(m)$ and $K_{\mathrm{fin}}(m)$ are the initial and final values of $K$ at the mass coordinate $m$; $\alpha$ ($0 \le \alpha \le 1$) is a parameter for interpolation and a surrogate for time. The correspondence between $\alpha$ and the actual time is not important for the current purpose. We map this $K$ profile for a given value of $\alpha$ onto the (spherical) reference configuration so that the mass coordinate should be preserved. We then calculate the rotational equilibrium configuration for the profile. The evolution via cooling is represented as an ensemble of these configurations for the values of $\alpha$ varying from $0$ to $1$ with the specific angular momentum fixed on each grid point in the Lagrangian coordinates.
This is certainly a very crude approximation. Since the main purpose of this calculation is to demonstrate the capability of the new formulation as a proof of principle, though, we think that the approximation is justified.

\section{Results}
\label{sec:result}

\subsection{Polytropes}

\begin{figure}
	\centering
	\includegraphics[width=\columnwidth]{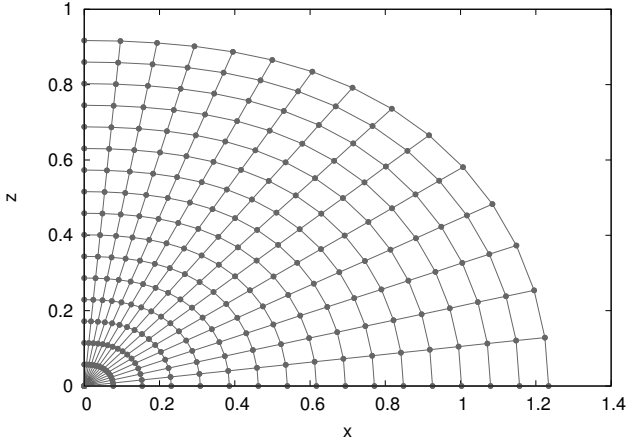}
	\caption{ A grid configuration after the equlibrium is reached with remapping for the model with the polytropic index $N=1.5$ and $(N_r,N_{\theta})=(16,16)$. The ratio of the centrifugal force to the gravitational force is 0.743 on the equatorial surface.}
	\label{fig:structureN1.5}
\end{figure}

We begin with the results for the polytrope with $N=1.5$ as a canonical case. We deploy a Lagrangian mesh with $(N_r, N_{\theta})=(16, 16)$.
Figure~\ref{fig:structureN1.5} shows that the mesh configuration after mechanical equilibrium is established for the rotating star with the total angular momentum $\hat{J_{\mathrm{tot}}}=0.182$ in the current normalization. We stress that this configuration is obtained from a spherical reference configuration by increasing the specific angular momentum at each grid point in the Lagrangian coordinates gradually by a common factor. Note that the initial profile of the specific angular momentum on the Lagrangian coordinates is given by equation~(\ref{eq:J_law}). At this value of $\hat{J_{\mathrm{tot}}}$, the ratio of the centrifugal force to the gravitational force on the equatorial surface is 0.213; the ratio of the rotational energy to the (absolute value of) gravitational energy is $T/|W|=5.46\times10^{-2}$; and the ratio of the polar radius to the equatorial radius is 0.743. In this calculation, the remapping is implemented. Unlike the calculation for Fig.~\ref{fig:distortN1.5}, in which the remapping was not done, the mesh is stretched smoothly in the radial direction.

\begin{figure}
	\centering
	\includegraphics[width=\columnwidth]{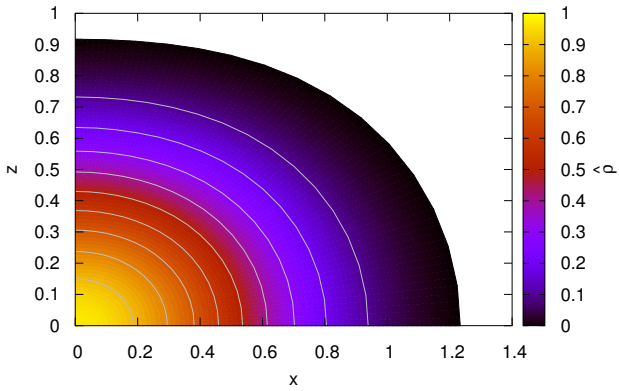}
	\caption{The colour map of density with some contour lines in the first quadrant of the meridian section for the same model as in Fig.~\ref{fig:structureN1.5}.}
	\label{fig:DenmapN1.5}
\end{figure}

\begin{figure}
	\centering
	\includegraphics[width=\columnwidth]{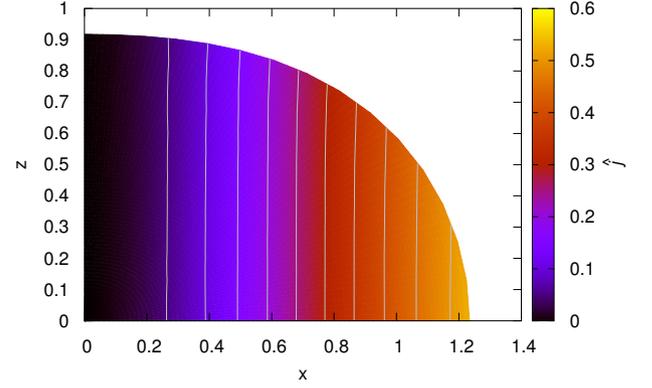}\\
    \includegraphics[width=\columnwidth]{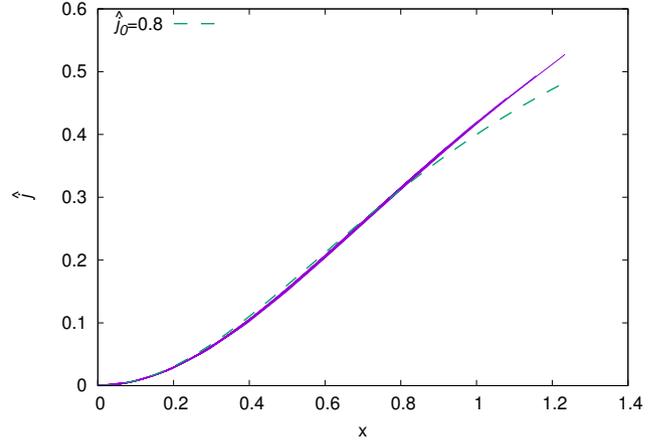}
	\caption{(Upper panel:) Same as Fig.~\ref{fig:structureN1.5} but for the specific angular momentum
	(Lower panel:) the specific angular momenta for all grid points as a function of the distance from the rotation axis $x=R\sin\Theta$. The points for the same radial Lagrangian coordinates are connected by a line. For reference, the curve shown in equation~(\ref{eq:J_law}) for $\hat{j_0}=0.8$ is plotted by the green dotted line.}
	\label{fig:AMmapN1.5}
\end{figure}

In Fig.\ref{fig:DenmapN1.5}, we exhibit the corresponding density distribution as a color map with some contour lines. Note that the density is obtained from the mass and volume of each cell, the latter of which is derived from the grid configuration calculated. It is found that isopycnic surfaces are also oblate in accordance with the motion of grid points.
In Fig.\ref{fig:AMmapN1.5}, we present the distribution of specific angular momentum for the same model: in the upper panel, it is shown as a color map with some contour lines in the meridian section; in the lower panel the specific angular momentum is plotted as a function of the distance from the rotation axis, $x=R\sin\Theta$, for all grid points. From the upper panel, it is already apparent that the rotation is cylindrical, a result expected for the barotropic case. This is even more evident from the lower panel, where all the points sit on a single curve, an implying the specific angular momentum is a function of $x=R \sin \Theta$ alone. Note that in this plot we connect by a line the points with the same radial Lagrangian coordinates; there are hence $N_r$ lines drawn but the difference is so small that they look like a single line. We stress again that this is a result of the grid relocation and is not trivial at all in our Lagrangian formulation. This is one of the clear indications that our calculation is correct. It is also mentioned that the spatial profile of specific angular momentum is different from that give initially in the reference configuration, equation~(\ref{eq:J_law}). For reference, the curve shown in equation~(\ref{eq:J_law}) for $\hat{j_0}=0.8$ is plotted by the green dotted line. It can be seen that it is not consistent with equation~(\ref{eq:J_law}), especially away from the axis.

\begin{figure}
	\centering
	\includegraphics[width=\columnwidth]{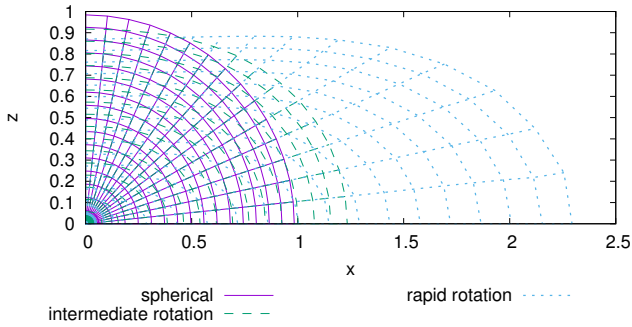}
	\caption{Comparison of the grid configurations for different rotation rates. The numbers of the grid points are $(N_r, N_{\theta})=(16,16)$ and the polytropic index $N=1.5$. The magenta solid, green dashed and cyan dotted curves correspond to the models with no, intermediate and rapidly rotations, respectively. The intermediate-rotation model is the same as for Fig.\ref{fig:structureN1.5}, whereas the rapid rotation model has the ratio of the centrifugal force to the gravitational force of 0.587 on the equatorial surface.}
	\label{fig:MeshN1.5}
\end{figure}

\begin{figure}
	\centering
	\includegraphics[width=\columnwidth]{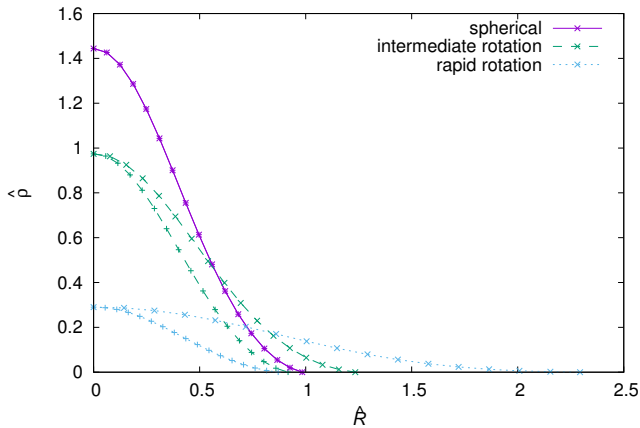}
	\caption{Comparison of the density profiles on the rotation axis (+) and on the equator (×) for the models with different rotation rates. The numbers of the grid points are $(N_r, N_{\theta})=(16,16)$ and the polytropic index is $N=1.5$. The colours of lines are the same as in Fig.\ref{fig:MeshN1.5}.}
	\label{fig:DensityN1.5}
\end{figure}

The grid configurations and the density distributions on the rotation axis and the equator for the same model but different rotation rates are displayed in Figs.~\ref{fig:MeshN1.5} and \ref{fig:DensityN1.5}, respectively.
The intermediate rotation model in these figures is the same as the model shown in Fig.~\ref{fig:structureN1.5}. The rapid rotation model has $\hat{J_{\mathrm{tot}}}=0.483$; the ratio of the centrifugal force to the gravitational force is $0.587$; $T/|W|=0.168$; the aspect ratio is $0.380$. As the rotational velocity increases, the grid is elongated in the equatorial direction and is shrunken slightly in the polar direction as demonstrated in Fig.~\ref{fig:MeshN1.5}. It should be obvious that it would be all but impossible to obtain such stretched grid configurations without remapping. In fact, \citet{Yasutake2016} were not able to deal with such a high deformation with their triangular mesh.
Given in Fig.\ref{fig:DensityN1.5} are the density distributions on the rotation axis and on the equator. As the rotation rate increases and the star becomes more oblate, the density decreases as a whole. Since the star is stretched in the equatorial direction, the density gradient is much smaller in this direction compared with that in the polar direction.

As a direct confirmation that our calculations are correct, we compare the results obtained with our Lagrangian formulation with those calculated with a well-calibrated Eulerian code of Fujisawa \citep{Fujisawa2015}. Extracting the spatial distributions of angular momentum and entropy, and feeding them together with total mass to the Eulerian code, we obtain the rotational equilibrium configuration on the Eulerian grid that corresponds to the original configuration on the Lagrangian grid. The rotational polytrope model here is the same as that in Fig.~\ref{fig:structureN1.5}. The numbers of the Eulerian grid points are $N_r=257$ and $N_{\theta}=129$, much larger than those for the Lagrangian code: $(N_r, N_{\theta})=(16, 16) - (32, 32)$. We may hence regard the Eulerian result as the exact solution.

\begin{figure}
	\centering
	\includegraphics[width=\columnwidth]{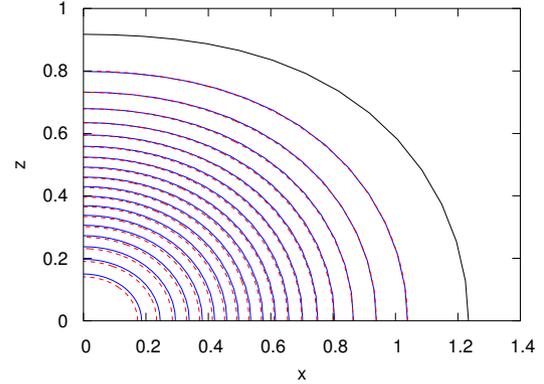}
	\caption{Comparison of the density distributions obtained with the Eulerian calculation (red dashed lines) and the Lagrangian calculation (blue solid lines) for the same spatial distribution of the angular momentum. The numbers of the grid points are $(N_r,N_{\theta})=(16,16)$ and the polytropic index is $N=1.5$. The outermost black solid curve represents the stellar surface of the Lagrange formulation.}
	\label{fig:Euler}
\end{figure}

It is clear from Fig.~\ref{fig:Euler} that the two results agree with each other fairly well. The relative deviation is of the order of $10^{-2}$ throughout the star even in the case of the lowest resolution model ($N_r=N_{\theta}=16$). This is in sharp contrast to the results of \citet{Yasutake2015,Yasutake2016}, who deployed the triangular Lagrangian grid and looked for a configuration with the minimum energy; they found relative errors in density of the order of unity near the stellar surface.

We also look at the Virial constant $V_c$ defined in equation~(\ref{eq:Virial}).
The values of $V_c$ for the current polytrope model are $V_c = 1.25\times 10^{-2}, 7.21\times 10^{-3}, 4.92\times 10^{-3}, $ for $(N_r, N_{\theta}) = (16,16)$, $(24,24)$, $(32,32)$, respectively, in the Lagrangian models. Reflecting the finer grid employed, the Eulerian calculation achieves $V_c = 1.97\times10^{-5}$.
The results for the Lagrangian calculations clearly show the numerical convergence of our models with the number of grid points (more on this point later in this section). It is also noted that \citet{Yasutake2016} obtained $V_c<10^{-3}$ typically with $\sim500$ triangular cells, more or less the same accuracy as ours in this measure, although they were not able to treat more rapidly rotating models.

\begin{figure}
	\centering
	\includegraphics[width=\columnwidth]{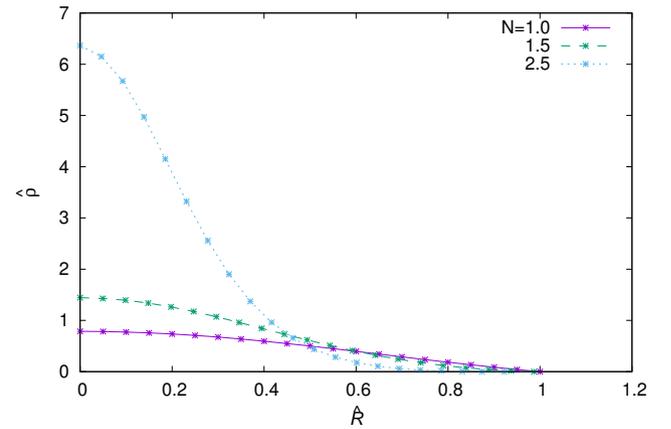}
	\caption{The density distributions of spherically symmetric polytropes with $N=1.0$ (magenta solid), $1.5$ (green dashed) and $2.5$ (cyan dotted).}
	\label{fig:DensityN}
\end{figure}

It is well known that the density distribution is sensitive to polytropic index $N$.
We show in Fig.~\ref{fig:DensityN} the density profiles for the non-rotating, spherically symmetric polytropes with $N=1.0, 1.5$ and $2.5$ for reference. One finds more centrally-concentrated configurations with larger values of $N$ as expected.
We compare the Virial constants for these models, both non-rotational and rotational, changing the number of radial grid points $N_r(=N_{\theta})$.
The rotational models have $\hat{J_{\mathrm{tot}}}=0.12, 0.11, 6.7\times 10^{-2}$, the ratios of the centrifugal force to the gravitational force on the equatorial surface are $8.23\times10^{-2}, 9.90\times10^{-2}, 9.65\times10^{-2}$, $T/|W|=2.65\times10^{-2}, 2.64\times10^{-2}, 1.26\times10^{-2}$ and the ratios of the polar radius to the equatorial radius are $0.879, 0.875, 0.887$ for $N=1.0, 1.5, 2.5$, respectively.

\begin{figure}
	\centering
	\includegraphics[width=\columnwidth]{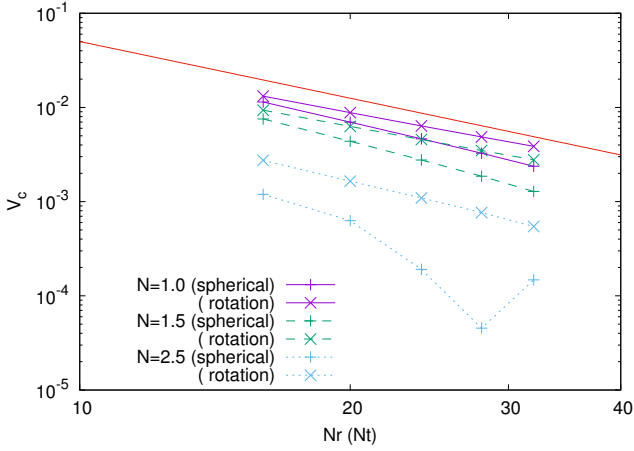}
	\caption{The Virial constants as a function of number of grid points for the rotational polytropes with different values of the polytropic index. The magenta solid, green dashed, and cyan dotted curves indicate the results for $N=1.0, 1.5$ and $2.5$, respectively. The orange line shows the slope of -2 for reference. The values of Virial constants are summarized in Table\ref{table:Virial}.}
	\label{fig:Virial2}
\end{figure}

The results are summarized in Table\ref{table:Virial}.
\begin{table*}
  \centering
  \caption{The Virial constants for the non-rotating and rotating models with the polytropic index $N=1.0, 1.5, 2.5$ and the numbers of grid points $(N_r, N_{\theta})=(16,16), (20,20), (24,24), (28,28)$. The definition of the Virial constant is given in equation~(\ref{eq:Virial}).}
  \begin{tabular}{ccccccccc}
    \hline
    \multirow{2}{*}{$N$} &  \multicolumn{2}{c}{$16\times16$} & \multicolumn{2}{c}{$20\times20$} & \multicolumn{2}{c}{$24\times24$} & \multicolumn{2}{c}{$28\times28$}  \\
    & spherical & rotation & spherical & rotation & spherical & rotation & spherical & rotation  \\
    \hline \hline
    $1.0$ & $1.14\times10^{-2}$ & $1.32\times10^{-2}$ & $6.99\times10^{-3}$ & $8.80\times10^{-3}$ & $4.65\times10^{-3}$ & $6.37\times10^{-3}$ & $3.27\times10^{-3}$ & $4.86\times10^{-3}$ \\
    $1.5$ & $7.54\times10^{-3}$ & $9.35\times10^{-3}$ & $4.37\times10^{-3}$ & $6.29\times10^{-3}$ & $2.76\times10^{-3}$ & $4.56\times10^{-3}$ & $1.86\times10^{-3}$ & $3.49\times10^{-3}$ \\
    $2.5$ & $1.19\times10^{-3}$ & $2.74\times10^{-3}$ & $6.27\times10^{-4}$ & $1.65\times10^{-3}$ & $1.91\times10^{-4}$ & $1.09\times10^{-3}$ & $4.54\times10^{-5}$ & $7.66\times10^{-4}$ \\
    \hline
  \end{tabular}
  \label{table:Virial}
\end{table*}
They are also plotted in Fig.~\ref{fig:Virial2}.
One finds that all cases show similar convergence behavior: $V_c \propto N_r^{-2}$, although the absolute values tend to be larger for the rotational models with smaller values of $N$. It is nice that the value of $V_c$ is smaller than $1\%$ for all cases. The quadratic dependence on $1/N_r$ indicates that our finite-difference scheme is of second-order accuracy, which is indeed as expected. The apparent higher accuracy in the result for the non-rotational model with $N=2.5$ at $N_r=28$ is accidental.

\subsection{Baroclinic Rotational Equilibria}

\begin{figure}
	\centering
	\includegraphics[width=\columnwidth]{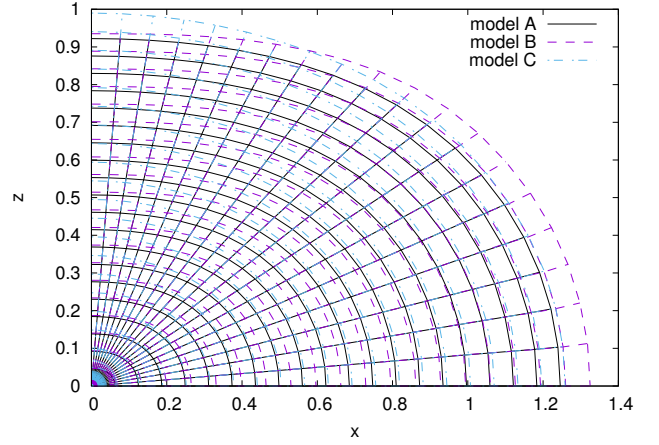}
	\caption{Comparison of the grid configurations for the baroclinic models in Table~.\ref{table:baroclinic}. The numbers of grid points are $(N_r,N_{\theta})=(20,20)$ and the index in the \ac{EOS} is $N=1.5$. The black solid, magenta dashed and cyan dotted curves denote the models A, B and C in Table~\ref{table:baroclinic}, respectively.}
	\label{fig:baroclinic}
\end{figure}

Now we proceed to the baroclinic models. The values of $\epsilon_1, \epsilon_2$ employed here are summarized in Table~\ref{table:baroclinic} together with the polar and equatorial radii obtained. The specific angular momentum profile on the Lagrangian grid is the same as that of the model shown in Fig.~\ref{fig:structureN1.5}.
\begin{table}
  \centering
  \caption{Summary of the baroclinic models. The same rotational rate is the same as in Fig.\ref{fig:structureN1.5}.}
  \begin{tabular}{ccccc}
    \hline
    Model  & $\epsilon_1$ & $\epsilon_2$ & $\hat{R}_{\mathrm{pol}}$ & $\hat{R}_{\mathrm{eq}}$  \\
    \hline \hline
    A & 0.0 & 0.0 & 0.922 & 1.246  \\
    B & 0.44 & 0.0 & 0.935 & 1.325 \\
    C & 0.0 & 0.44 & 0.990 & 1.259 \\
    \hline
  \end{tabular}
  \label{table:baroclinic}
\end{table}
The configurations for the three models with $N=1.5$ are compared in Fig.~\ref{fig:baroclinic}. The number of grid points are $(N_r, N_{\theta})=(20, 20)$ this time. The black solid curves show the result for model A, which takes $\epsilon_1=\epsilon_2=0$ and is actually barotropic, whereas the magenta dashed curves are for model B with $(\epsilon_1, \epsilon_2)=(0.44,0)$ and the cyan dot-dashed curves represent model C with $(\epsilon_1, \epsilon_2)=(0,0.44)$.
As expected from the functional form in equation~(\ref{eq:baroclinic}), a positive $\epsilon_1$ tends to expand the star in the equatorial direction while a positive $\epsilon_2$ is inclined to bloat it in the polar direction.

\begin{figure*}
	\begin{minipage}[b]{0.45\linewidth}
		\centering
		\includegraphics[width=\columnwidth]{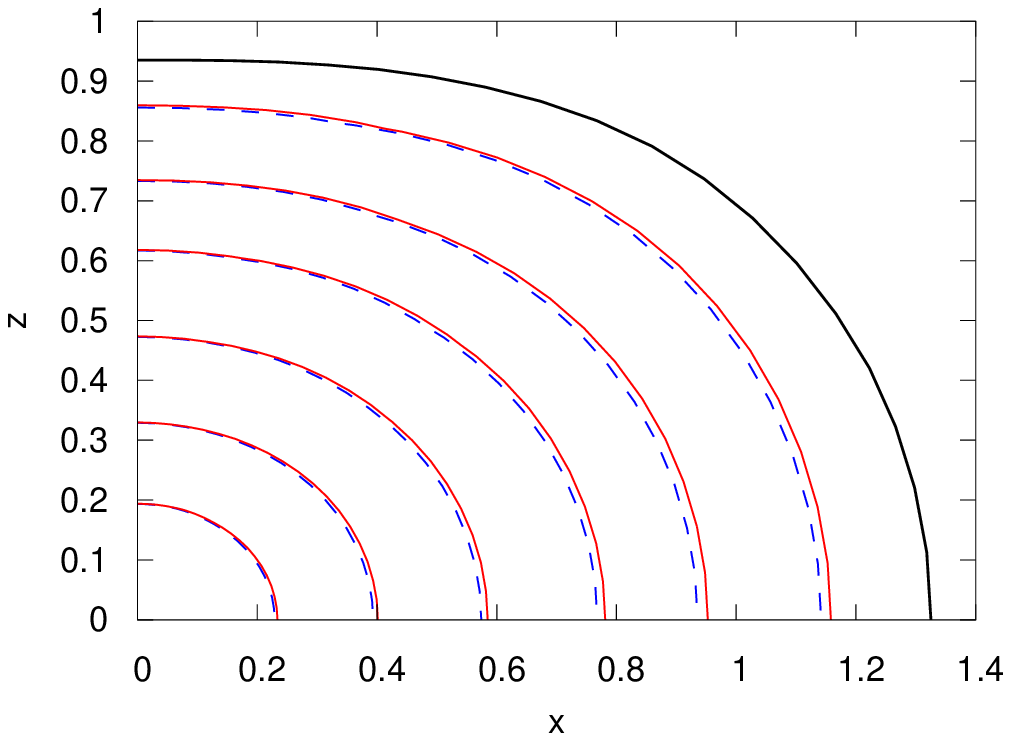}
		\subcaption{Density (model B)}
	\end{minipage}
	\begin{minipage}[b]{0.45\linewidth}
		\centering
		\includegraphics[width=\columnwidth]{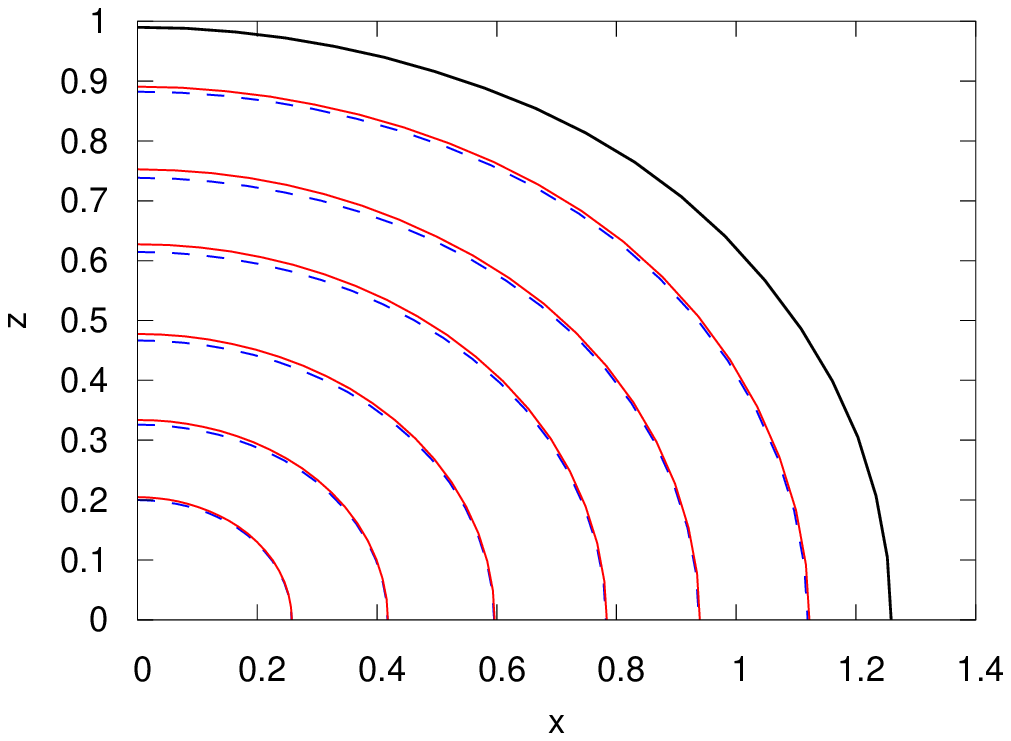}
		\subcaption{Density (model C)}
	\end{minipage}\\
	\begin{minipage}[b]{0.45\linewidth}
		\centering
		\includegraphics[width=\columnwidth]{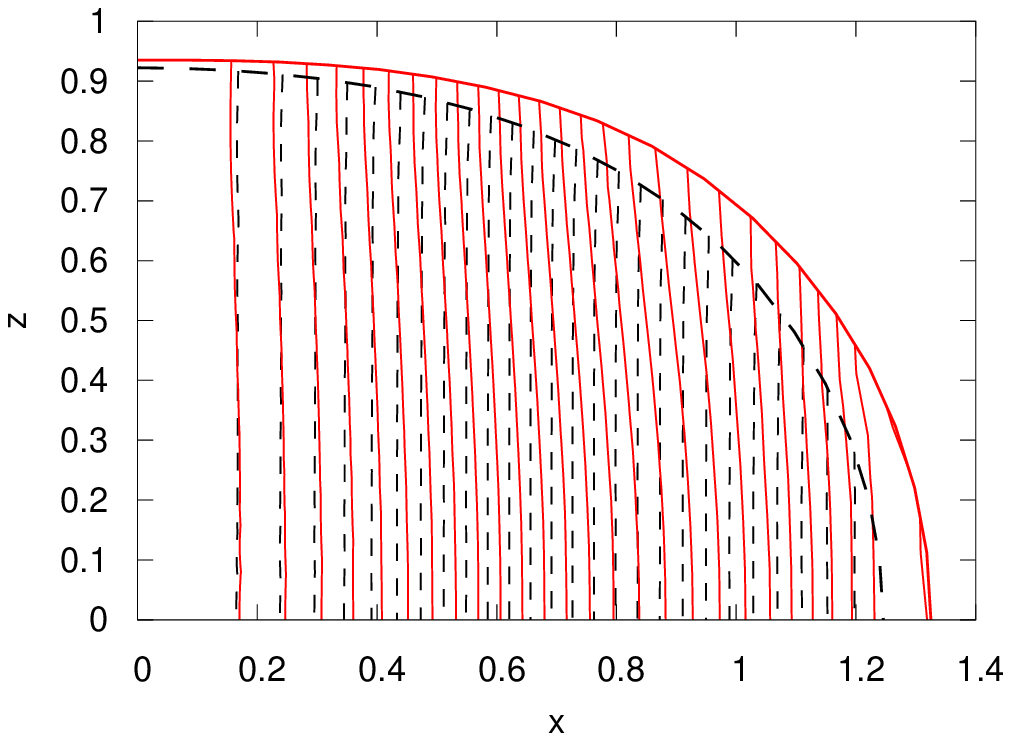}
		\subcaption{Specific Angular Momentum (model B)}
	\end{minipage}
	\begin{minipage}[b]{0.45\linewidth}
		\centering
		\includegraphics[width=\columnwidth]{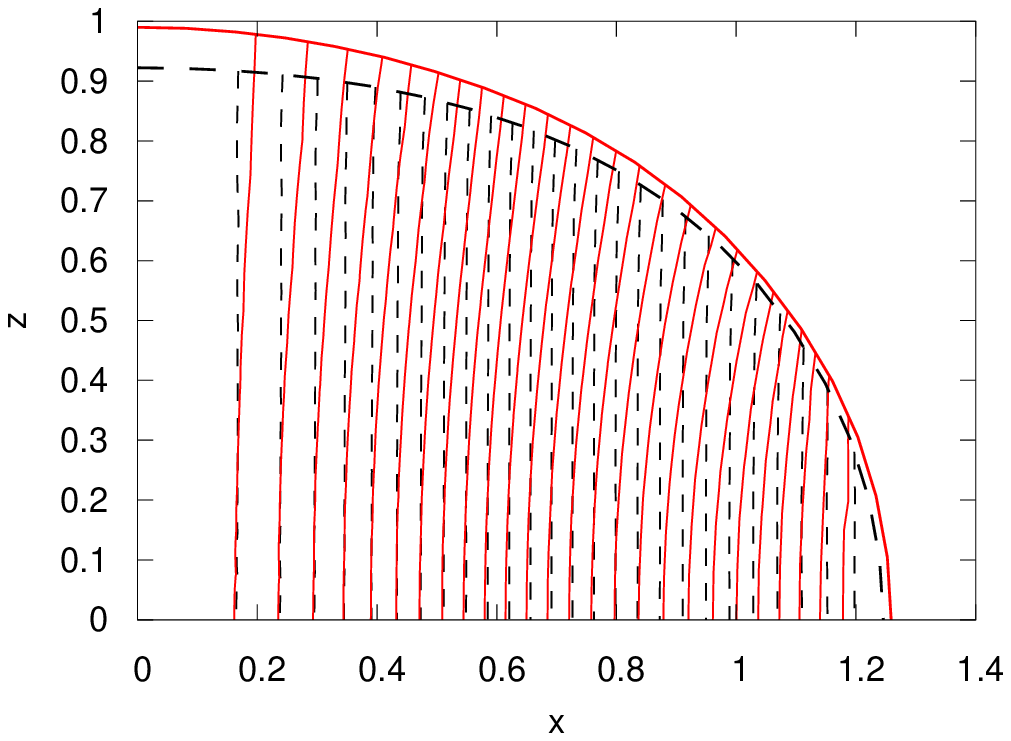}
		\subcaption{Specific Angular Momentum (model C)}
	\end{minipage}\\
	\begin{minipage}[b]{0.45\linewidth}
		\centering
		\includegraphics[width=\columnwidth]{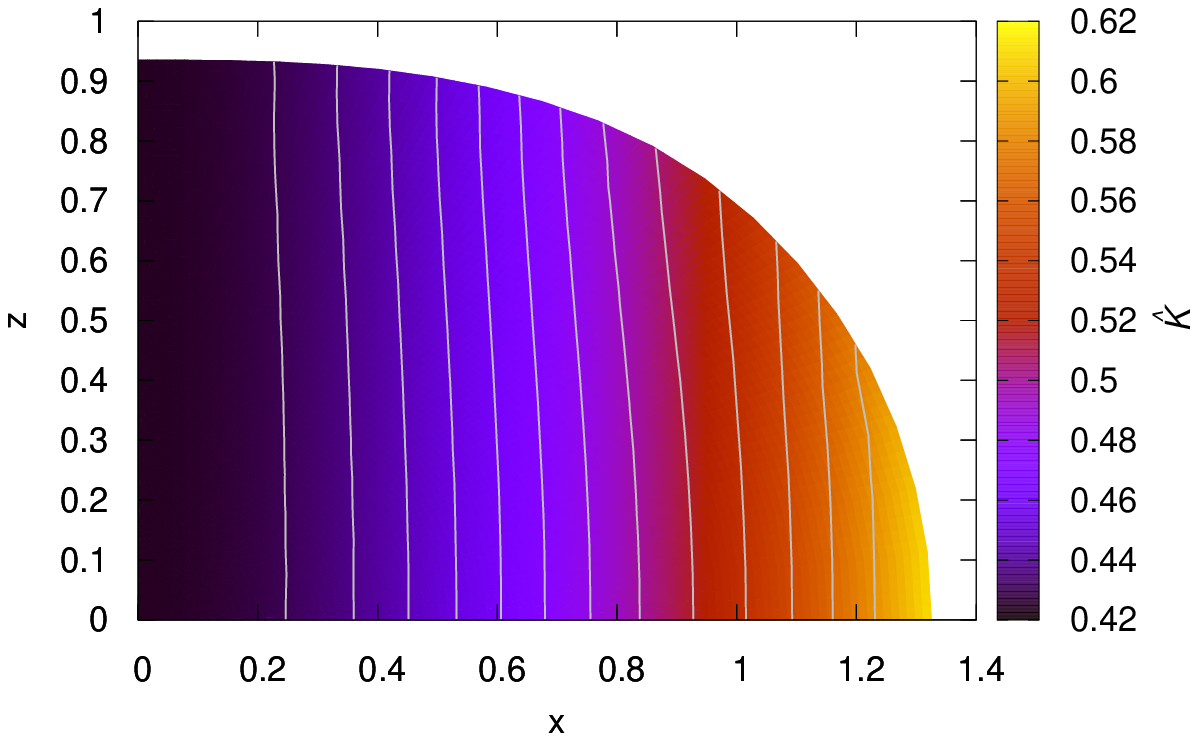}
		\subcaption{Specific Angular Momentum and K (model B)}
	\end{minipage}
	\begin{minipage}[b]{0.45\linewidth}
		\centering
		\includegraphics[width=\columnwidth]{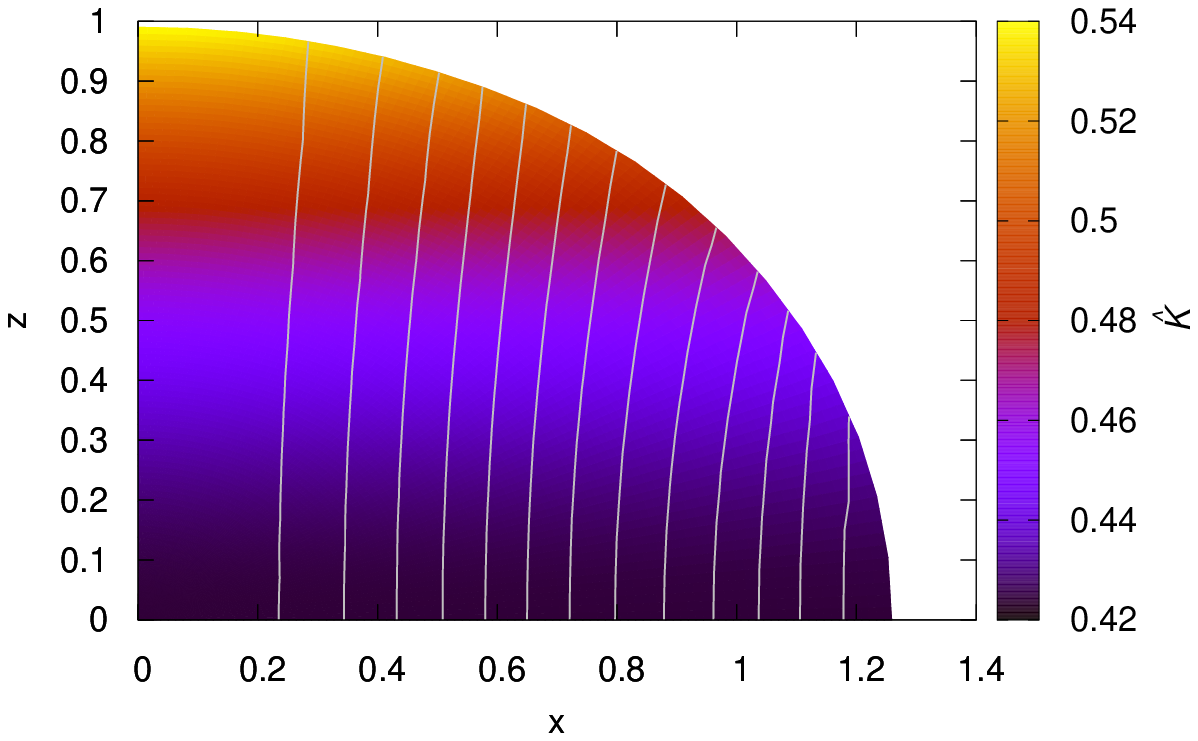}
		\subcaption{Specific Angular Momentum and K (model C)}
	\end{minipage}
	\caption{Comparison of the baroclinic models B (left panels) and C (right panels).
    (Upper panels:) The density and pressure distributions in the first quadrant of the meridian section are shown as contour plots. The blue dashed and red solid curves represent the density and pressure, respectively.
	(Middle panels:) The distributions of the specific angular momentum are presented as color maps. The red solid curves exhibit the results of model B or C. For reference, the black dashed curves indicate the result for model A. In both panels, the stellar surfaces are indicated by thicker curves.
	(Bottom panels:) The distributions of the specific angular momentum are shown with contour lines whereas those of K are presented in color.}
	\label{fig:baroclinic_all}
\end{figure*}

In Fig.\ref{fig:baroclinic_all}, we display the distributions of the density, the pressure, the specific angular momentum and the coefficient $K$ as contours in the (first quadrant of) meridian section for the three models.
The upper two panels exhibit the density and pressure for models B (left) and C (right); the blue dashed curves and the red solid curves correspond to the density and pressure, respectively. In both panels the density contours are inclined against pressure contours, a clear indication of baroclinicity. It can be seen that the isobaric curves are more oblate than the isopycnic ones in model B while the opposite is true in model C.
The middle two panels present the distributions of the specific angular momentum. The red solid curves show the results for model B (C) in the left (right) panel, and the black dashed curves give the result for model A for comparison. One finds that the iso-specific-angular-momentum lines of model A are parallel to the rotation axis, whereas those of model B are leaned to the axis, i.e., $dj/dz>0$ and hence $d\Omega/dz>0$. The trend in model C is opposite.
According to the Bjerknes-Rosseland rule, $d\Omega/dz>0$ implies that the isobaric surfaces are more oblate than the isopycnic surfaces and, as a result, the temperature on the isobaric surface is lower toward the poles \citep{Tassoul1978}. We confirm that our results are in agreement with this rule.

We mention the dynamical stability of our baroclinic rotational models here.
By the Solberg-H\o iland criterion, a baroclinic rotational star is dynamically stable if and only if the following two conditions are satisfied simultaneously: (i) the specific entropy rises outward (more precisely, in the opposite direction to the pressure gradient), (ii) on all isentropic surfaces, the specific angular momentum increases from the pole toward the equator \citep{Tassoul1978}.
The bottom two panels in Fig.\ref{fig:baroclinic_all} show the specific angular momentum as contours and the coefficient $K$, the surrogate for the specific entropy, as color maps, respectively. It is clear that the specific entropy never decreases outward. One finds, on the other hand, that the specific angular momentum increases from pole to equator on the isentropic surfaces for model C, whereas for model B the opposite holds. This means that model C is dynamically stable but model B is unstable.
Our formulation can produce both stable and unstable configurations, the fact that will be useful to study the hydrodynamical instability numerically by using these models as the unperturbed states.

Note in passing that the values of the Virial constant are $8.81\times 10^{-3}$ and $9.13\times 10^{-3}$ for models B and C, respectively. They are more or less the same as the barotropic counterparts.

\subsection{Cooling of a Rotational WD}

\begin{figure*}
	\begin{minipage}[b]{0.45\linewidth}
		\centering
		\includegraphics[width=\columnwidth]{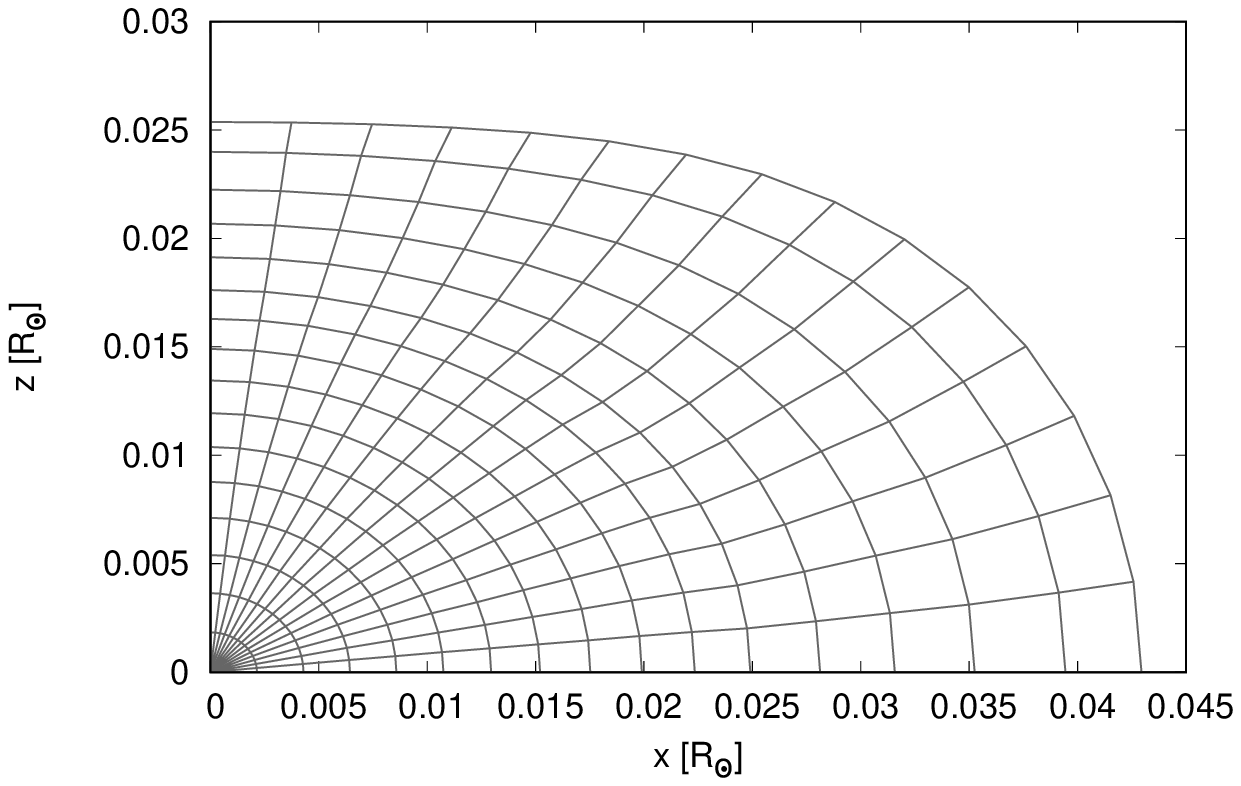}
		\subcaption{Structure (Initial)}
	\end{minipage}
	\begin{minipage}[b]{0.45\linewidth}
		\centering
		\includegraphics[width=\columnwidth]{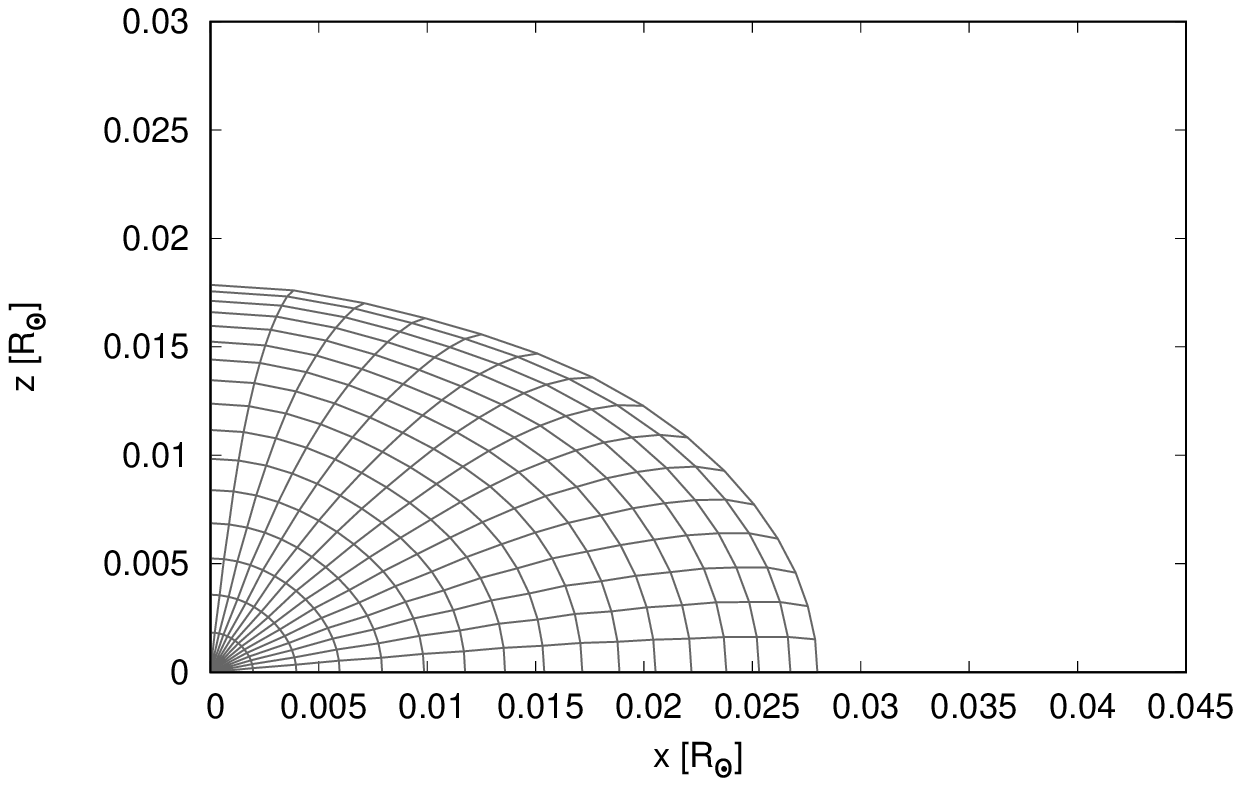}
		\subcaption{Structure (After Cooling)}
	\end{minipage}\\
	\begin{minipage}[b]{0.45\linewidth}
		\centering
		\includegraphics[width=\columnwidth]{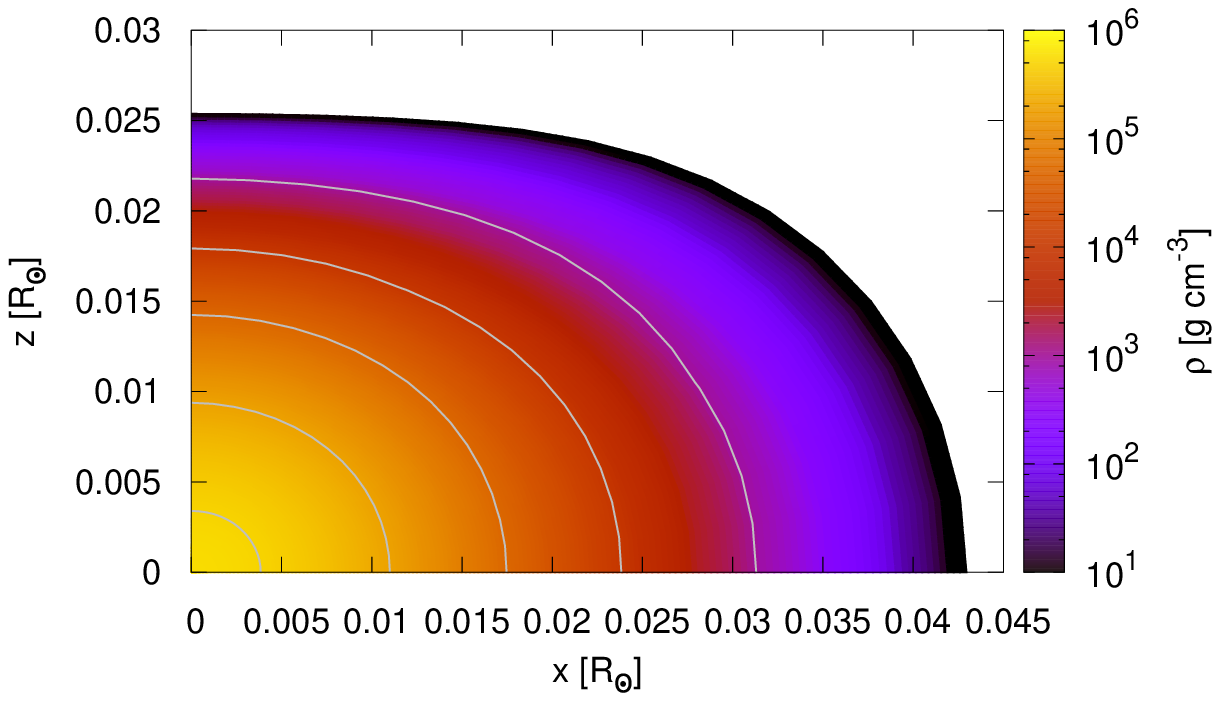}
		\subcaption{Density (Initial)}
	\end{minipage}
	\begin{minipage}[b]{0.45\linewidth}
		\centering
		\includegraphics[width=\columnwidth]{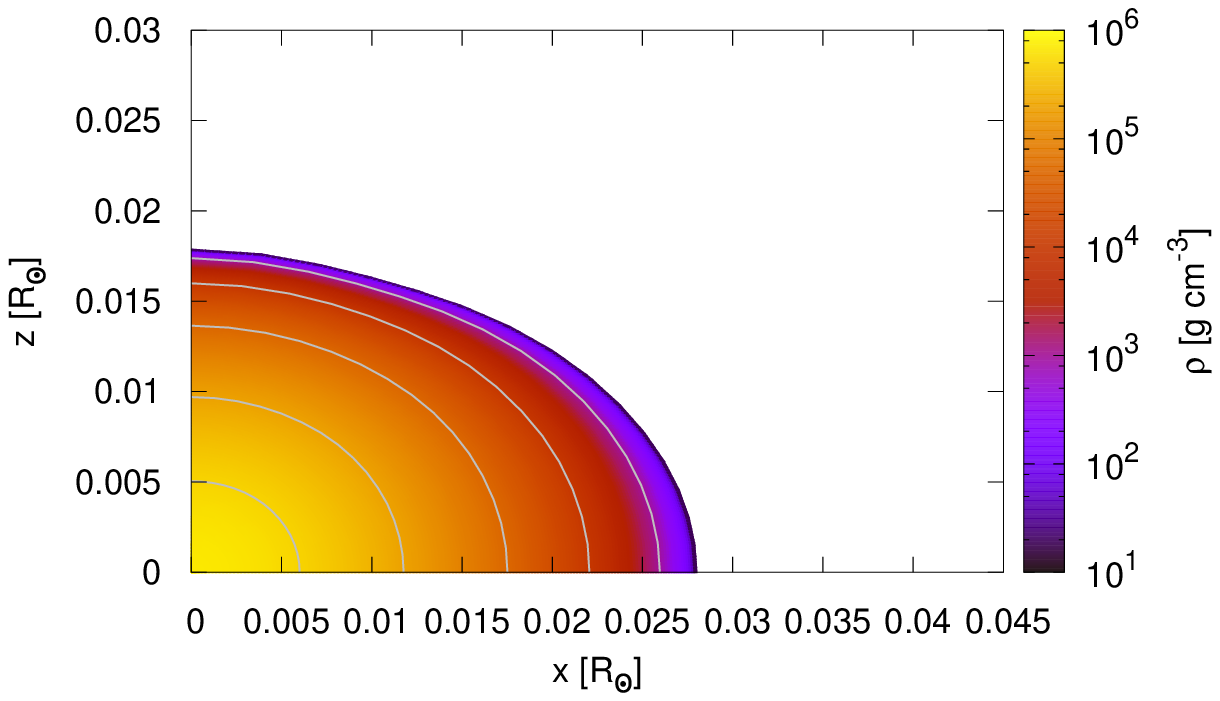}
		\subcaption{Density (After Cooling)}
	\end{minipage}\\
	\begin{minipage}[b]{0.45\linewidth}
		\centering
		\includegraphics[width=\columnwidth]{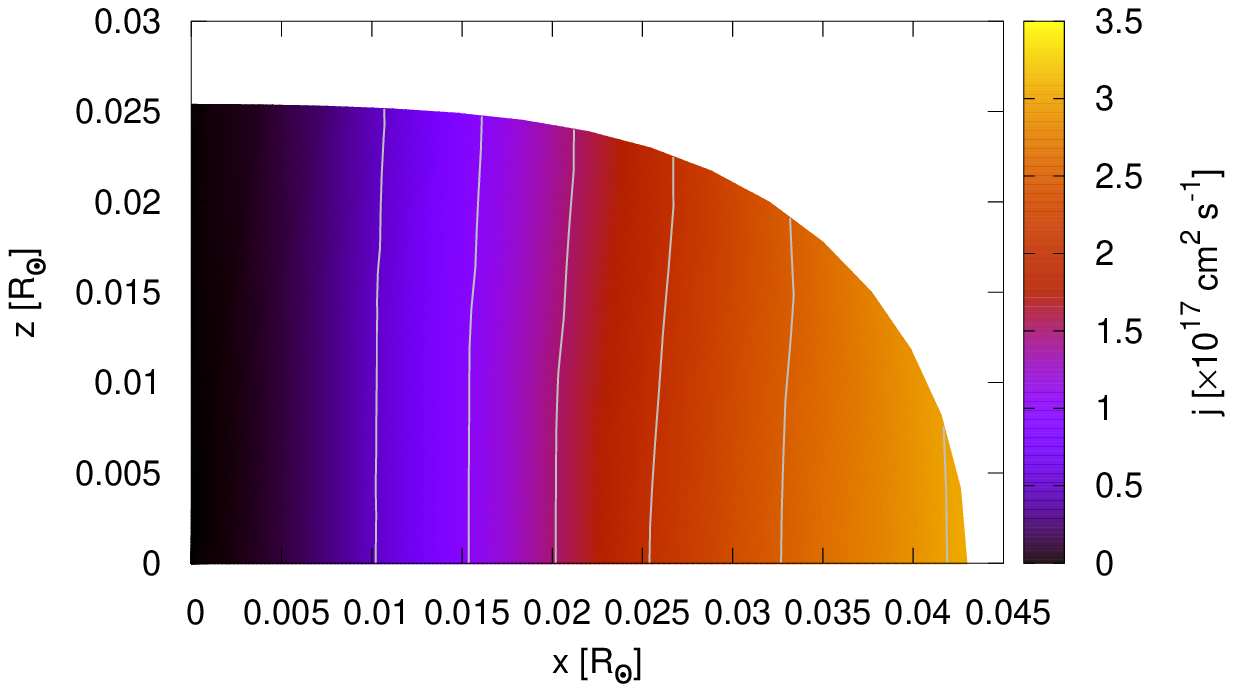}
		\subcaption{Specific Angular Momentum (Initial)}
	\end{minipage}
	\begin{minipage}[b]{0.45\linewidth}
		\centering
		\includegraphics[width=\columnwidth]{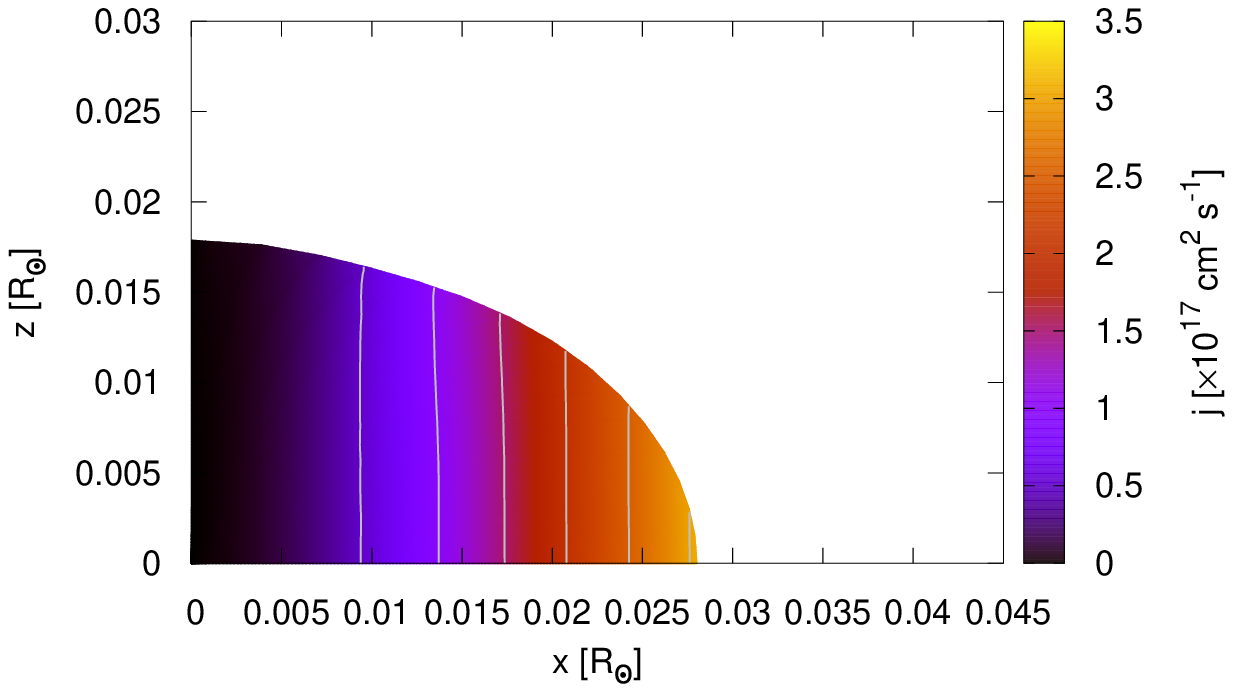}
		\subcaption{Specific Angular Momentum (After Cooling)}
	\end{minipage}
	\caption{Results of the \ac{WD} cooling calculation. The numbers of the grid points are $(N_r,N_{\theta})=(16,16)$ and the index in the \ac{EOS} is $N=2.5$. Various quantities before and after cooling are exhibited in the left and right panels, respectively. (Top panels:) the grid configurations are shown. Note that the regridding is turned off in this calculation. (Middle panels:) the density distributions are displayed as color maps. (Bottom panels:) the distributions of the specific angular momentum are shown as color maps with some contour lines.
	}
	\label{fig:WD}
\end{figure*}

The results of the toy model calculation of \ac{WD} cooling are given here. In Fig.\ref{fig:WD}, we show the grid configurations, the density distributions and the specific angular momentum distributions at the initial ($\alpha=0$) and final ($\alpha=1$) times. We can see in the upper panels that the \ac{WD} shrinks as it cools, with its total angular momentum $J_{\mathrm{tot}}=7.34\times10^{49}~\mathrm{erg~s}$ preserved. As a result, the \ac{WD} spins faster and become more flattened.
At this value of $J$, the ratios of the centrifugal force to the gravitational force on the equatorial surface are $0.384, 0.574$ before and after cooling, respectively; the ratios of the rotational energy to the gravitational energy are $T/|W|= 5.15\times10^{-2}, 5.86\times10^{-2}$; and the ratios of the polar radius to the equatorial radius are $0.591, 0.638$ and the Virial constants $V_c$ are $9.63\times10^{-3}, 1.46\times10^{-2}$. In this calculation, the remapping is not performed on purpose, so that one could see clearly how the fluid elements move as \ac{WD} cools and is spun up. It is indeed evident that they are all relocated significantly toward the equator particularly near the surface.
The middle panels show that the density increases as a whole as it contracts.
Furthermore, the distribution of specific angular momentum shown in the bottom panels are not cylindrical especially at the initial time, since $K$ is not constant. As the cooling proceeds, however, it becomes more cylindrical. We stress again that the advantage of the Lagrangian formulation is that the evolution of the angular momentum distribution as a result of the change in the stellar configuration can be derived automatically.

\section{Conclusions}
\label{sec:conclusion}

We have proposed a new formulation to obtain numerically the equilibrium structure of a rapidly rotating star on the two-dimensional Lagrangian coordinates. We assign a trio of the mass, the specific angular momentum and the specific entropy to the fluid elements, or the Lagrangian grid points in the discretized version, initially, and their positions that could satisfy the force-balance equations and the Poisson equation for self-gravity simultaneously. For the numerical solution of the force-balance equations, we have used the W4 method, a new multi-dimensional iterative root-finding scheme of our own devising, which could be an alternative to the Newton-Raphson method and has a global convergence \citep[][]{Okawa2023}. We have also employed a spectral method for the Poisson equation. Furthermore, we have implemented the remapping procedure in the formulation so that too much mesh deformation, which is a major issue in the Lagrangian formulation, degrading accuracy and sometimes hampering the convergence of the iterative solution of the equations, could be avoided. We have conducted some test calculations to demonstrate the performance and capability of our new formulation.

For the polytropic models we have shown first that our formulation can treat highly non-spherical configurations unlike our previous attempt with the triangular mesh based on the variation principle \citep{Yasutake2015,Yasutake2016}. Utilizing the well-calibrated Eulerian code of \citet{Fujisawa2015}, we have made a quantitative comparison with the Eulerian solution with the same spatial distributions of density and angular momentum. We have seen that the Lagrangian results agree well with the Eulerian counterpart even near the stellar surface. Varying the polytropic index, we have observed that the Virial constant becomes smaller quadratically with the number of grid points for all cases. It is noteworthy that the rotation is found to be cylindrical as should be for the barotropic \ac{EOS}, since this is not at all trivial for the Lagrangian formulation, in which the spatial distribution of the specific angular momentum is determined as a result of the relocation of all fluid elements.

In the baroclinic models, in which the \ac{EOS} is not a function of density alone but depends on other thermodynamic variables, e.g., the specific entropy as in our cases, we have observed that the isobaric surfaces do not coincide with the isopycnic surfaces indeed.
Moreover, we have confirmed that they satisfy the Bjerknes-Rosseland rule. The rotation is no longer cylindrical and we have also seen that the inclination of the iso-specific-angular-momentum surfaces is consistent with that of the isobaric surfaces against the isopycnic surfaces.

We have touched the Solberg-H\o iland criterion for the dynamical instability in the rotational equilibrium configurations. We have found that one of the baroclinic configurations we constructed is stable but the other is unstable. Note that even the construction of unstable configurations is useful. Indeed, they can be employed as the unperturbed states in the numerical study of the instability.

In the toy model calculation for the cooling of a rotational \ac{WD}, we have constructed a series of rotational equilibrium configurations, which have the same profile of specific angular momentum on the Lagrangian coordinates but have consecutively smaller specific entropies. In so doing, we have mimicked the evolution of entropy in the one-dimensional MESA simulation for the same \ac{WD}. We have demonstrated that our new formulation successfully yields a contraction of the \ac{WD} accompanied by a spin-up and, as a consequence, a flattening. We have observed that the rotation becomes more cylindrical as the cooling proceeds.

These results are encouraging but are admittedly very crude approximations to reality. We have omitted possible fluid motions in the meridian section such as convection and circulation. The energy generation and transfer should be incorporated for the application to stellar evolution. We need to consider possible angular momentum transport between fluid elements. Magnetic fields should be implemented on the same basis. These are major issues on our agenda list and will be addressed one after another in the forthcoming publications. Last but not least we are also developing the general relativistic formulation with the application to rotating relativistic objects such as neutron stars \citep{Okawa2022}.

\section*{Acknowledgements}
This work was supported by JSPS KAKENHI Grant Number 20K03951, 20K03953, 20K14512, 20H04728, 20H04742.
%%%%%%%%%%%%%%%%%%%%%%%%%%%%%%%%%%%%%%%%%%%%%%%%%%
\section*{Data Availability}

The data underlying this paper will be available from the corresponding author on reasonable request.



\bibliographystyle{mnras}
\bibliography{example} % if your bibtex file is called example.bib

%\begin{thebibliography}{99}
%\end{thebibliography}

%%%%%%%%%%%%%%%%%%%%%%%%%%%%%%%%%%%%%%%%%%%%%%%%%%

%%%%%%%%%%%%%%%%% APPENDICES %%%%%%%%%%%%%%%%%%%%%

\appendix

\section{Some known properties of rotational equilibria}
\label{app:barotrope}

The distribution of angular velocity (and hence of specific angular momentum as well) in the rotational equilibrium configuration is known to be cylindrical in the barotropic case. This is obtained as follows: the Euler equations on the cylindrical coordinates ($\varpi, \varphi, z$) are written as
\begin{align}
    \frac{1}{\rho}\frac{\partial P}{\partial\varpi} &= -\frac{\partial\phi}{\partial\varpi}+\Omega^2\varpi, \\
    \frac{1}{\rho}\frac{\partial P}{\partial z}     &= -\frac{\partial\phi}{\partial z},
\end{align}
where permanent rotation is assumed. By eliminating the gravitational potential $\phi$ from these equations, the following relationship is obtained:
\begin{align}
    2\frac{\partial\Omega}{\partial z}{\bm v}_{\mathrm{\varphi}} = \mathrm{grad}\frac{1}{\rho}\times\mathrm{grad}P
\end{align}
where $\bm{v}_{\mathrm{\varphi}}$ represents the rotational velocity.
In the barotropic case, $P=P(\rho)$,
\begin{equation}
    \mathrm{grad}\frac{1}{\rho}\times\mathrm{grad}P=0 \ \ \ \mathrm{i.e.}\ \ \  \frac{\partial\Omega}{\partial z}=0.
\end{equation}
This implies that the angular velocity depends only on the distance from the axis, i.e., the rotation is cylindrical.

On the other hand, in the baroclinic case, this is not true, and the iso-angular-velocity surfaces (and hence the iso-specific-angular-momentum surfaces as well) are no longer parallel to the rotation axis.
Indeed, we find $\partial\Omega/\partial z\gtrless0 \leftrightarrow \mathrm{grad}\rho\times\mathrm{grad}P\lessgtr0$, i.e., the inclination of the isopycnic surface against the isobaric surface is determined by the signature of $\partial \Omega / \partial z$. In addition, if $T\propto P/\rho$, the temperature on an isobaric surface is lower (higher) toward the poles if $\partial \Omega / \partial z$ is positive (negative). These are called the Bjerkness-Rosseland rule.

\section{Virial relation}
\label{app:Virial}

The Virial theorem is derived from the following volume integral of the Euler equation:
\begin{align}
    \int {\bm r} \cdot \bm\nabla P dV + \int {\bm r} \cdot \rho \bm\nabla \phi dV - \int {\bm r} \cdot \frac{1}{2}\rho\Omega^2\nabla(R\sin\Theta)^2 dV = 0
\end{align}
The integration by parts of the first term gives
\begin{align}
    \int {\bm r} \cdot \bm\nabla P dV = - 3 \int P dV \equiv -3 U.
\end{align}
The second term, on the other hand, becomes
\begin{align}
    \int {\bm r} \cdot \rho({\bf r}) \bm\nabla \phi dV
  &= -G \int \int \rho({\bf r}) \rho({\bf r}')
      \frac{{\bf r}\cdot({\bf r}-{\bf r}')}{|{\bf r}-{\bf r}'|^3} dV'
      dV \notag \\
  &= -\frac{1}{2}G \int \int
      \rho({\bf r})\rho({\bf r}')\frac{({\bf r} -
      {\bf r}')^2}{|{\bf r} - {\bf r}'|^3}dV' dV \notag \\
  &= -\frac{1}{2}\int \rho({\bf r})\phi({\bf r}) \, dV \equiv -W.
\end{align}
The third term is calculated as
\begin{align}
    -\frac{1}{2} \int {\bm r} \cdot \rho\Omega^2\nabla(R\sin\Theta)^2 dV
    = -\int \rho \Omega^2(R\sin\Theta)^2 dV    \equiv -2T.
\end{align}
Finally we obtain the Virial relation as
\begin{align}
    3U  + W + 2T = 0.
\end{align}

%%%%%%%%%%%%%%%%%%%%%%%%%%%%%%%%%%%%%%%%%%%%%%%%%%

% Don't change these lines
\bsp	% typesetting comment
\label{lastpage}
\end{document}